\documentclass{aastex}
\usepackage{spr-astr-addons}
\usepackage{url}

\RequirePackage{color}

\newcommand{\emaila}{jameel@giki.edu.pk}
\usepackage{latexsym}
\usepackage{graphics}
\usepackage{graphicx}
\usepackage{epsfig}

\begin{document}

\title{Study of Gamow-Teller transitions in isotopes of titanium within the quasi particle random phase approximation}

\shorttitle{Study of GT transitions in isotopes of Ti in the QRPA }

\author{Sadiye Cakmak \altaffilmark{1,2}, Jameel-Un Nabi \altaffilmark{2},
Tahsin Babacan  \altaffilmark{1}  and Cevad Selam \altaffilmark{3} }

\altaffiltext{1}{Department of Physics, Celal Bayar University,
Manisa, Turkey}

\email{\emaila} \altaffiltext{2}{Faculty of Engineering Sciences,
GIK Institute of Engineering Sciences and Technology, Topi 23640,
Swabi, Khyber Pakhtunkhwa, Pakistan}

\altaffiltext{3}{Department of Physics, Alparslan University, Mus,
Turkey}
 \altaffiltext{} {This study is supported by BAP project of
Celal Bayar University with number 2013-004}
\begin{abstract}
The Gamow-Teller (GT) transition is inarguably one of the most
important nuclear weak transitions of the spin-isosopin $\sigma\tau$
type. It has many applications in nuclear and astrophysics. These
include, but are not limited to, r-process $\beta$-decays, stellar
electron captures, neutrino cooling rates, neutrino absorption and
inelastic scattering on nuclei. The quasiparticle random phase
approximation (QRPA) is an efficient way to generate GT strength
distribution. In order to better understand both theoretical
systematics and uncertainties, we compare the GT strength
distributions, centroid and width calculations for $^{40-60}$Ti
isotopes, using the pn-QRPA, Pyatov method (PM) and the Schematic
model (SM). The pn-QRPA and SM are further sub-divided into three
categories in order to highlight the role of particle-particle (pp)
force and deformation of the nucleus in the GT strength
calculations. In PM, we study only the influence of the pp force in
the calculation. We also compare with experimental results and other
calculations where available. We found that the inclusion of pp
force and deformation significantly improves the performance of SM
and pn-QRPA models. Incorporation of pp force leads to pinning down
the centroid value in the PM. The calculated GT strength functions
using the pn-QRPA (C) and SM (C) models are in reasonable agreement
with measured data.
\end{abstract}


\section{Introduction}
The Gamow-Teller (GT) response of nuclei in the medium mass region
are crucial prerequisites in order to determine the precollapse
evolution of a supernova \cite{Ost92}. GT excitations act only on
the spin-isospin ($\sigma\tau$) degrees of freedom. The isospin
operator in spherical coordinates has three components
$\tau_{\pm,0}$. Here, the plus sign refers to Gamow-Teller
(GT$_{+}$) transitions in which a proton is changed into a neutron
(e.g. in $\beta^{+}$ decays and electron captures), while the minus
sign corresponds to GT$_{-}$ transitions in which a neutron is
changed into a proton (realized in $\beta^{-}$-decays). The total
GT$_{-}$  and GT$_{+}$ strengths, noted as $S_{-}$ and $S_{+}$,
respectively, are connected by the model independent Ikeda sum rule
as $S_{-} - S_{+} = 3(N -Z)$, where $N$ and $Z$ are the numbers of
neutrons and protons \cite{Ike63}. The third component GT$_{0}$ is
of relevance to inelastic neutrino-nucleus scattering for low
neutrino energies and it would not be further considered in this
manuscript. The GT transitions in $fp$-shell nuclei play decisive
roles in presupernova phases of massive stars and also during the
core collapse stages of supernovae, specially in neutrino induced
processes. The lepton fraction ($Y_{e}$) of the stellar matter is
one of the factors that controls the gravitational core-collapse of
massive stars. It is the degeneracy pressure of the leptons which
counters the mammoth gravitational force of massive stars. Once the
lepton contents of the stellar matter reduce, the core collapses
within fractions of a second. The lepton content of the stellar
matter in turn is governed by $\beta$-decay and electron capture
rates among iron-regime nuclides. The $\beta$-decay gives positive
contribution whereas electron capture rates give a negative
contribution to $Y_{e}$. The time evolution of $Y_{e}$ is a crucial
parameter which is also a key to generate a successful explosion in
modeling and simulation of core-collapse supernovae. Nuclei in the
mass range A $\sim$ 60, at stellar densities less than around
10$^{11}$ gcm$^{-3}$, posses electron chemical potential of the same
order of magnitude as the nuclear Q-value. Under such conditions the
electron capture rates are sensitive to the detailed GT
distributions.  A reliable and microscopic calculation of ground and
excited states GT distribution functions is then in order. At much
higher stellar densities, the electron chemical potential is much
larger than Q-values. For high densities electron capture rates are
more sensitive to the total GT strength. Additionally for higher
densities centroids and widths of the GT distribution become
important parameters for estimation of weak-interaction rates. Hence
one needs not only the microscopic GT strength distribution
functions but also the total strength, centroid and width of the
distributions to reliably calculate electron capture rates in
stellar matter under given physical conditions.

GT distributions have been extracted experimentally using different
techniques. The isovector response of nuclei may be studied using
the nucleon charge-exchange reactions $(p,n)$ (e.g. \cite{And91}) or
$(n,p)$ (e.g. \cite{Kat94}); by other reactions such as
$(^{3}$He,$t$) (e.g. \cite{Fuj99}) or $($t$,^{3}$He) (e.g.
\cite{Cole06}), ($d,^{2}$He) (e.g. \cite{Bau05}) or through heavy
ion reactions (e.g. \cite{Ber97}). The GT cross sections ($\Delta T
=1, \Delta S =1, \Delta L =0, \hspace{0.1cm} 0\hbar\omega$
excitations) are proportional to the analogous beta-decay strengths
at vanishing linear momentum transfer. Charge-exchange reactions at
small momentum transfer can therefore be used to study beta-decay
strength distributions when beta-decay is not energetically
possible. The $(p,n)$, $(^{3}$He,$t)$ reactions probe the GT$_{-}$
strength whereas the $(n,p)$, $(d,^{2}$He) reactions give the
GT$_{+}$ strength.

One also requires GT strength distributions of hundreds of unstable
nuclei which requires much effort and technology to be studied
experimentally. The situation is improving with the construction of
next-generation radioactive ion-beam facilities. To further
complicate the matters, one also requires excited state GT strength
distribution functions in astrophysical environments (as parent
excited states are thermally populated at high stellar temperatures)
where no measured data is available. Consequently, astrophysical
calculations rely heavily on detailed microscopic calculations.

Theoretical calculations of GT transitions fall generally into three
major categories: simple independent-particle models; full-scale
interacting shell-model calculations; and, in between, the
random-phase approximation (RPA) and quasi-particle random-phase
approximation (QRPA). The simple independent-particle models have
reported to pose problems in correct placement of  GT centroid.
These models place the centroid of the GT strength too high for
even-even parent nuclides and too low on odd-A and odd-odd parents
\cite{Pin00}. Full interacting shell-model calculations are
computationally demanding although one can exploit the Lanczos
algorithm to efficiently generate the strength distribution
\cite{SMreview05}. For medium-mass nuclei, one still needs to choose
from among a number of competing semi-realistic/semi-empirical
interactions in shell model calculations. RPA and QRPA can be
treated as approximations to a full shell-model calculation and they
are much less demanding computationally. Furthermore, they have the
additional advantage that one can employ a separable multi-shell
interaction which in turn grant access to a huge model space, up to
$7\hbar\omega$, to perform the calculations.

The main difficulty with both experiment and theory is that the
strength distribution connects to many states. In contrast to Fermi
transitions, which relate a parent state to a single daughter state
(the isobaric analog state), GT transitions are fragmented over many
daughter states. This is caused by the fact that the GT operator
does not commute with the residual interaction beyond the mean-field
approximation which gives rise to shell model single-particle
orbits.

Three widely used model within the QRPA formalism are the pn-QRPA
model, the Schematic Model (SM) and the Pyatov Method (PM). The
pn-QRPA theory is an efficient way to generate GT strength
distributions.  It was Halbleib and Sorenson \cite{Hal67} who
generalized this model to describe charge-changing transitions of
the type $(Z,N) \rightarrow (Z \pm 1, N \mp 1)$ whereas the usual
RPA was formulated for excitations in the same nucleus. The model
was then extended from spherical to deformed nuclei (using
Nilsson-model wave functions) by Krumlinde and M\"{o}ller
\cite{Kru84}. Further extension of the model to treat odd-odd nuclei
and transitions from nuclear excited states was done by Muto and
collaborators \cite{Mut92}. It was Nabi and Klapdor-Kleingrothaus
who used the pn-QRPA theory for the first time to calculate the
stellar weak interaction rates over a wide range of temperature and
density scale for sd- \cite{Nab99} and fp/fpg-shell nuclei
\cite{Nab04} in stellar matter (see also Ref. \cite{Nab99a}). Since
then, these calculations were further refined with use of more
efficient algorithms, computing power, incorporation of latest data
from mass compilations and experimental values, and fine-tuning of
model parameters (e.g. see
\cite{Nab05,Nab07,Nab08b,Nab09,Nab10,Nab11,Nab12,Nab13}). There is a
considerable amount of uncertainty involved in all types of
calculations of stellar weak interactions. The uncertainty
associated with the microscopic calculation of the pn-QRPA model was
discussed in detail in Ref. \cite{Nab08b}. The reliability of the
pn-QRPA calculations was discussed in length by Nabi and
Klapdor-Kleingrothaus \cite{Nab04}. There, the authors compared the
measured data (half lives and B(GT$_{\pm}$) strength) of thousands
of nuclide with the pn-QRPA calculations and got fairly good
comparison.

The formalism developed by Pyatov \cite{Pya77} has been applied to
different problems for more than 35 years. The main step of Pyatov
formalism is based on the definition of the effective Hamiltonian
which considers Dirac restrictions \cite{Dirac64}. This formalism
has been used in problems which include the violation of the
particle number \cite{Civ88}, invariant invariance \cite{Civ90},
generalized Galielo invariance \cite{Civ92} and velocity dependent
effect \cite{Sak91}. Moreover, this formalism has also been applied
by Civitarese et. al. \cite{Civ98} to the isospin dependent
Hamiltonians in quasi-particle basis in order to investigate the
relation between the collapse of RPA solutions and the violation of
isospin symmetry. It has also been used by Magierski and Wyss
\cite{Mag00}. In studies done by Kuliev and collaborators
\cite{Kul00} and Babacan et. al. \cite{Bab04,Bab05,Bab05a}, this
method was applied to scissor mode vibrations, isobar analog states
(IAS) and isospin admixtures in the ground states of spherical
nuclei as well as Gamow-Teller Resonance (GTR) states. Coherent
excitation of the nucleons results in concentration of most of total
GT transition strength in a narrow excitation region in the daughter
nucleus known as GTR states. These resonances correspond to the
coherent proton-hole proton and neutron-hole neutron excitations in
the known electromagnetic giant resonances. Schematic Model is a
special case of Pyatov method and it would be discussed further in
the next section.

Weak rates on titanium isotopes have numerous astrophysical
applications. The estimated $^{44}$Ti yield from post explosive
nucleosynthesis supernova debris can be given as an example and it
can be used to test and calibrate the supernova models \cite{Sar99}.
$\beta$-decay of $^{40,41}$Ti has implications for solar-neutrino
detection \cite{Liu98}. The large energy release of the $^{40}$Ti
decay [$Q_{ec}$ = 11680(160) KeV] enables one to extract all
information that is relevant for the GT contribution to the rate of
solar neutrinos absorbed by $^{40}$Ar. Titanium isotopes have been
assigned different nucleosynthetic origins. Oxygen burning for
$^{46}$Ti, silicon and carbon burning for $^{47,49}$Ti, silicon
burning for the abundant isotope $^{48}$Ti, and carbon burning for
$^{50}$Ti. The difference in these origins helps us to explain the
life time of the Galaxy and its gradual evolution \cite{Nab07}.
Aufderheide and collaborators \cite{Auf94} searched for key weak
interaction nuclei in presupernova evolution. Phases of evolution in
massive stars, after core silicon burning, were considered and a
search was performed for the most important electron captures and
$\beta$-decay nuclei in these scenarios.  From these lists, electron
captures on $^{49,51,52,53,54}$Ti and $\beta$-decay of
$^{51,52,53,54,55,56}$Ti were short-listed to be of astrophysical
importance. Again, per simulation results of Heger and collaborators
\cite{Heg01}, weak rates on isotopes of titanium are considered as
very important for presupernova evolution of massive stars.

In this paper, we calculate and compare GT strength distributions of
isotopes of titanium (mass number 40 to 60) using three different
(microscopic) QRPA models namely pn-QRPA, Pyatov Method (PM) and the
Schematic Model (SM). All three models are further divided into two sub-categories:\\
(A) Nuclei are treated as spherical and only particle-hole (ph)
interaction is taken into account. The resulting model is referred to as Model (A) in this manuscript.\\
(B)  The particle-particle interaction is usually thought to play a
minor role in $\beta^{-}$ decay but has shown to be of decisive
importance in $\beta^{+}$ decay and electron capture reactions
\cite{Hir93} in pn-QRPA models. Particle-particle (pp) interaction
is then incorporated into the Model (A) to get our Model (B).\\

The pn-QRPA and SM are further classified into a third category in
order to study the effect of deformation of nucleus on the GT
strength functions. The deformation parameter is recently argued to
be one of the most important parameters in pn-QRPA calculations
\cite{Ste04}. In order to check the effect of incorporation of
deformation in the pn-QRPA and SM models, we finally lift the
initial assumption of spherical nuclei and introduce deformations in
the Model (B) to get Model (C). In other words, Model (C) takes into
account nuclear deformations and also perform calculation in
\textit{both} pp \textit{and} ph channels. Incorporation of
deformation into PM is currently being worked on and it would be
taken up as a future assignment.

The next section briefly describes the theoretical formalism used to
calculate the GT strength distributions using the PM, SM and the
pn-QRPA theory. The calculated GT$_{\pm}$ strength distributions for
titanium isotopes are presented and compared with measurements and
against other theoretical calculations in Sec. 3.  The main
conclusions of this work are finally presented in Sec. 4.

\section{Theoretical Formalism}
\subsection{Pyatov Method and the Schematic Model}
Restoration of the broken super symmetry property in static pairing
interaction potential is of great importance for GT transitions. On
the other hand, no effect of the translational and rotational
invariance violations on GT transitions is seen. The Pyatov Method
(PM) used in this paper provides this restoration in two different
ways: addition of static interaction potential to the total
Hamiltonian after its broken symmetry property has been restored and
the restoration of the broken symmetry in quasi particle space. We
employ the second method in this article. The Schematic Model (SM)
is a special case of the PM where we exclude the effective
interaction term from the total Hamiltonian. We next consider the
case of even-even and odd-A cases separately and describe the
necessary formalism within the PM and SM.

\subsubsection{Even-Even Nuclei}
The Schematic Model (SM) Hamiltonian for GT excitations in the quasi
particle representation is given as
\begin{eqnarray}
H_{SM}=H_{SQP}+h_{ph}+h_{pp}, \label{Eqt. 1}
\end{eqnarray}
where $H_{SQP}$ is the Single Quasi Particle (SQP) Hamiltonian,
$h_{ph}^{GT}$ and $h_{pp}^{GT}$ are the GT effective interactions in
the ph and pp channels, respectively \cite{Nec10}. The effective
interaction constants in the ph and pp channel were fixed from the
experimental value of the Gamow-Teller resonance (GTR) energy and
the $\beta^{±}$-decay \textit{logft} values between the low energy
states of the parent and daughter nucleus, respectively. In order to
restore the super symmetry property of the pairing part in total
Hamiltonian, certain terms which do not commute with the GT operator
were excluded from total Hamiltonian and the broken commutativity of
the remaining part due to the shell model mean field approximation
was restored by adding an effective interaction term $h_{0}$
\begin{eqnarray}
[H_{SM}-h_{ph}^{GT}-h_{pp}^{GT}-V_{1}-V_{c}-V_{ls}+h_{0},G_{1\mu}^{\pm}]=0,
\label{Eqt. 2}
\end{eqnarray}
or
\begin{eqnarray}
[H_{SQP}-V_{1}-V_{c}-V_{ls}+h_{0},G_{1\mu}^{\pm}]=0, \label{Eqt. 3}
\end{eqnarray}
where $V_{1}$, $V_{c}$ and $V_{ls}$ are isovectors, Coulomb and spin
orbital term of the shell model potential, respectively. The
restoration term $h_{0}$ in Eq.~(\ref{Eqt. 3}) is included in a
separable form:
\begin{eqnarray}
h_{0}=\sum_{\rho=\pm}\frac{1}{2\gamma_{\rho}}\sum_{\mu=0,\pm1}[H_{sqp}-V_{c}-V_{ls}-V_{1},G_{1\mu}^{\rho}]^{\dagger}\cdot\nonumber
\end{eqnarray}
\begin{eqnarray}
[H_{sqp}-V_{c}-V_{ls}-V_{1},G_{1\mu}^{\rho}] \label{Eqt. 4}.
\end{eqnarray}
The strength parameter $\gamma_{\rho}$ of $h_{0}$ effective
interaction is found from the commutation condition in
Eq.~(\ref{Eqt. 3}) and the following expression is obtained for this
constant (for details see Ref. \cite{Sal06}).
\begin{eqnarray}
\gamma_{\rho}=\frac{\rho}{2}\langle0|[[H_{sqp}-V_{c}-V_{ls}-V_{1},G_{1\mu}^{\rho}],G_{1\mu}^{\rho}]
|0\rangle.\nonumber
\end{eqnarray}
The total Hamiltonian of the system according to PM finally becomes
\begin{eqnarray}
H_{PM}=H_{SQP}+h_{0}+h_{ph}+h_{pp} \label{Eqt. 5}.
\end{eqnarray}

The eigenvalues and eigenfunctions of Hamiltonian given in
Eq.~(\ref{Eqt. 5}) are solved within the framework of the pn-QRPA
method. We considered the GT $1^{+}$ excitations in odd-odd nuclei
generated from the correlated ground state of the parent nucleus by
the charge-exchange spin-spin forces and used the eigenstates of the
single quasi particle Hamiltonian $H_{SQP}$ as a basis.

In pn-QRPA, the $i^{th}$ excited GT $1^{+}$ states in odd-odd nuclei
are considered as the phonon excitations and described by
\begin{eqnarray}
|1_{i}^{+}>=Q^{\dag}_{i}(\mu)|0>\nonumber
\end{eqnarray}
\begin{eqnarray}
=\sum_{np}[\psi_{np}^{i}C_{np}^{\dag}(\mu)-(-1)^{1+\mu}\varphi_{np}^{i}C_{np}(-\mu)]|0>
\label{Eqt. 6},
\end{eqnarray}
where $Q^{\dag}_{i}(\mu)$ is the pn-QRPA phonon creation operator,
$|0>$ is the phonon vacuum which corresponds to the ground state of
an even-even nucleus and fulfills $Q_{i}(\mu)|0>=0$ for all i. The
$\psi_{np}^{i}$ and $\varphi_{np}^{i}$ are quasi boson amplitudes.

Assuming that the phonon operators obey the commutation relations
\begin{eqnarray}
<0|[Q_{i}(\mu),Q^{\dag}_{j}(\mu')]|0>=\delta_{ij}\delta_{\mu\mu'},\nonumber
\end{eqnarray}
we obtain the following orthonormalization condition for amplitudes
$\psi_{np}^{i}$ and $\varphi_{np}^{i}$
\begin{eqnarray}
\sum_{np}[\psi_{np}^{i}\psi_{np}^{i'}-\varphi_{np}^{i}\varphi_{np}^{i'}]=\delta_{ii'}\label{Eqt.
7}.
\end{eqnarray}
The energies and wave functions of the GT $1^{+}$ states are
obtained from the pn-QRPA equation of motion:
\begin{eqnarray}
[H_{PM},Q^{\dag}_{i}(\mu)]|0>=\omega_{i}Q^{\dag}_{i}(\mu)|0>,
\label{Eqt. 8}
\end{eqnarray}
where $\omega_{i}$ is the energy of the GT $1^{+}$ states occurring
in neighboring odd-odd nuclei. We obtain the secular equation for
excitation energies $\omega_{i}$ of the GT $1^{+}$ states in the
neighbor odd-odd nuclei:
\begin{eqnarray}
[\chi_{+}-\sum_{np}\frac{\varepsilon_{np}(E_{np}^{(+)})^{2}}{\varepsilon_{np}^{2}-\omega_{i}^{2}}][\chi_{-}-\sum_{np}\frac{\varepsilon_{np}(E_{np}^{(-)})^{2}}{\varepsilon_{np}^{2}-\omega_{i}^{2}}]\nonumber
\end{eqnarray}
\begin{eqnarray}
-\omega_{i}^{2}[\sum_{np}\frac{E_{np}^{(+)}E_{np}^{(-)}}{\varepsilon_{np}^{2}-\omega_{i}^{2}}]^{2}=0
\end{eqnarray}
One of the characteristic quantities for the GT $1^{+}$ states
occurring in neighboring odd-odd nuclei is the GT transition matrix
elements. The $0^{+}\longrightarrow1^{+}$ $\beta^{-}$ and
$\beta^{+}$ transition matrix elements are calculated as
\begin{eqnarray}
M_{\beta^{-}}^{i}(0^{+}\longrightarrow1^{+}_{i})=<1_{i}^{+},\mu|G_{1\mu}^{-}|0^{+}>\nonumber
\end{eqnarray}
\begin{eqnarray}
=<0|[Q_{i}(\mu),G_{1\mu}^{-}]|0>\nonumber
\end{eqnarray}
\begin{eqnarray}
M_{\beta^{-}}^{i}(0^{+}\longrightarrow1^{+}_{i})=-\sum_{np}(\psi_{np}^{i}b_{np}+\varphi_{np}^{i}\bar{b}_{np}),
\label{Eqt. 9}
\end{eqnarray}
\begin{eqnarray}
M_{\beta^{+}}^{i}(0^{+}\longrightarrow1^{+}_{i})=<1_{i}^{+},\mu|G_{1\mu}^{+}|0^{+}>\nonumber
\end{eqnarray}
\begin{eqnarray}
=<0|[Q_{i}(\mu),G_{1\mu}^{+}]|0>\nonumber
\end{eqnarray}
\begin{eqnarray}
M_{\beta^{+}}^{i}(0^{+}\longrightarrow1^{+}_{i})=\sum_{np}(\psi_{np}^{i}\bar{b}_{np}+\varphi_{np}^{i}b_{np}).
\label{Eqt. 10}
\end{eqnarray}
$b_{np}$ and $\bar{b}_{np}$ are reduced matrix elements and they are
given by
\begin{eqnarray}
b_{np}=\frac{1}{\sqrt{3}}u_{j_{n}}v_{j_{p}}<j_{n}\parallel\sigma\parallel
j_{p}>\nonumber
\end{eqnarray}
\begin{eqnarray}
\bar{b}_{np}=\frac{1}{\sqrt{3}}u_{j_{p}}v_{j_{n}}<j_{n}\parallel\sigma\parallel
j_{p}>
\end{eqnarray}
where $v_{j_{n}}(u_{j_{n}})$ is the occupation (unoccupation)
amplitude which is obtained in the BCS calculations. The
$\beta^{\pm}$ reduced matrix elements are given by:
\begin{eqnarray}
B_{GT}^{(\pm)}(\omega_{i})=\sum_{\mu}|M_{\beta^{\pm}}^{i}(0^{+}\longrightarrow1^{+}_{i}|^{2}.
\label{Eqt. 11}
\end{eqnarray}
The $\beta^{\pm}$ transition strengths ($S^{\pm}$) defined as
\begin{eqnarray}
S^{\pm}=\sum_{i}B_{GT}^{(\pm)}(\omega_{i}), \label{Eqt. 12}
\end{eqnarray}
should fulfill the Ikeda Sum Rule (ISR)
\begin{eqnarray}
ISR=S^{(-)}-S^{(+)}\cong3(N-Z). \label{Eqt. 13}
\end{eqnarray}

\subsubsection{Odd-A Nuclei}
In pn-QRPA, wave function of the $i^{th}$ excited state in odd-A
nuclei is given by
\begin{equation}
|\psi_{I_{n}K_{n}}^{j}\rangle=\Omega_{I_{n}K_{n}}^{j^{\dag}}|0\rangle=[N_{I_{n}}^{j}\alpha_{I_{n}K_{n}}^{\dag}\nonumber
\end{equation}
\begin{equation}
+\sum_{iI_{p}K_{p}}R_{ij}^{I_{n}I_{p}}(I_{p}K_{p}1K_{n}-K_{p}/I_{n}K_{n})Q_{i}^{\dag}\alpha_{I_{p}K_{p}}^{\dag}]|0\rangle,
\label{Eqt. 14}
\end{equation}
where I is the total angular momentum and K is the projection of I
on the nuclear symmetry axis. It is assumed that wave functions for
odd-A nuclei is formed by superposition of the one quasi-particle,
and three quasi-particle (one quasi-particle + phonon) states. The
mixing amplitudes $N_{I_{n}^{j}}$ and $R_{ij}^{I_{n}I_{p}}$ are
fulfilled by the normalization condition
\begin{equation}
(N_{I_{n}}^{j})^{2}+\sum_{iI_{p}}(R_{ij}^{I_{n}I_{p}})^{2}=1.
\label{Eqt. 17}
\end{equation}
The energies and wave functions of the odd-A nuclei are obtained
from the pn-QRPA equation of motion:
\begin{equation}
[H,\Omega_{I_{n}K_{n}}^{j^{\dag}}]|0\rangle=W_{I_{n}K_{n}}^{j}\Omega_{I_{n}K_{n}}^{j^{\dag}}|0\rangle.
\label{Eqt. 18}
\end{equation}
the dispersion equation for excitation energies
$W_{I_{n}K_{n}}^{j}$, corresponding to states given in Eq. (16), is
obtained as
\begin{equation}
W_{I_{n}K_{n}}^{j}-E_{I_{n}K_{n}}=\nonumber
\end{equation}
\begin{equation}
2\sum_{i,I_{p},K_{p}}\frac{[X_{GT}^{ph}(d_{I_{n}I_{p}}M_{i}^{+}+\overline{d}_{I_{n}I_{p}}M_{i}^{-})-X_{GT}^{pp}(b_{I_{n}I_{p}}F_{i}^{+}+\overline{b}_{I_{n}I_{p}}F_{i}^{-})]^{2}}{W_{I_{n}K_{n}}^{j}-w_{i}-E_{I_{p}K_{p}}}
\end{equation}
Where $E_{I_{n}K_{n}}$ and $E_{I_{p}K_{p}}$ are neutron and proton
single quasi particle energies.
 The amplitude for three quasi-particle state
state,$R_{ij}^{I_{n}I_{p}}$, is written in terms of the amplitude
for one quasi-particle state, $N_{I_{n}}^{j}$, as follows:
\begin{equation}
R_{ij}^{I_{n}I_{p}}=\nonumber
\end{equation}
\begin{equation}
\frac{\sqrt{2}[X_{GT}^{ph}(d_{I_{n}I_{p}}M_{i}^{+}+\overline{d}_{I_{n}I_{p}}M_{i}^{-})-X_{GT}^{pp}(b_{I_{n}I_{p}}F_{i}^{+}+\overline{b}_{I_{n}I_{p}}F_{i}^{-})]}{W_{I_{n}K_{n}}^{j}-w_{i}-E_{I_{p}K_{p}}}N_{I_{n}}^{j}
\end{equation}
where $N_{I_{n}}^{j}$ is calculated from Eq. (17). The corresponding
expressions for the nuclei with odd-proton number are formulated by
performing the transformation $I_{n}K_{n}\leftrightarrow I_{p}K_{p}$
in Eqs. (16)-(20).
 The GT transition matrix elements of odd-A nuclei is
given by
\begin{equation}
M_{\beta^{\pm}}=\langle\psi_{I_{1}K_{1}}^{f}|\beta_{\mu}^{\pm}|\psi_{I_{2}K_{2}}^{i}\rangle.
\label{Eqt. 17}
\end{equation}
The corresponding matrix elements of odd-A transitions are expressed
for two different cases as follows: \\

(a) The case in which the number of pair does not change:
\begin{equation}
M_{\beta^{-}}=\langle\psi_{I_{p}K_{p}}^{f}|\beta_{\mu}^{-}|\psi_{I_{n}K_{n}}^{i}\rangle=\nonumber
\end{equation}
\begin{equation}
-[d_{I_{n}I_{p}}N_{I_{n}}^{i}N_{I_{p}}^{f}+\bar{d}_{I_{n}I_{p}}\sum_{j}R_{ij}^{I_{n}I_{p}}R_{fj}^{I_{n}I_{p}}\nonumber
\end{equation}
\begin{equation}
+N_{I_{n}}^{i}\sum_{j}R_{fj}^{I_{n}I_{p}}M_{j}^{-}+N_{I_{p}}^{f}\sum_{j}R_{ij}^{I_{n}I_{p}}M_{j}^{+}].
\label{Eqt. 18}
\end{equation}
\begin{equation}
M_{\beta^{+}}=\langle\psi_{I_{n}K_{n}}^{f}|\beta_{\mu}^{+}|\psi_{I_{p}K_{p}}^{i}\rangle=\nonumber
\end{equation}
\begin{equation}
-[d_{I_{n}I_{p}}N_{I_{n}}^{i}N_{I_{p}}^{f}+\bar{d}_{I_{n}I_{p}}\sum_{j}R_{ij}^{I_{n}I_{p}}R_{fj}^{I_{n}I_{p}}\nonumber
\end{equation}
\begin{equation}
+N_{I_{p}}^{i}\sum_{j}R_{fj}^{I_{n}I_{p}}M_{j}^{+}+N_{I_{n}}^{f}\sum_{j}R_{ij}^{I_{n}I_{p}}M_{j}^{-}.
\label{Eqt. 19}
\end{equation}

(b) The case in which the number of pair changes:
\begin{equation}
M_{\beta^{-}}=\langle\psi_{I_{n}K_{n}}^{f}|\beta_{\mu}^{-}|\psi_{I_{p}K_{p}}^{i}\rangle\nonumber
\end{equation}
\begin{equation}
=-[\bar{d}_{I_{n}I_{p}}N_{I_{p}}^{i}N_{I_{n}}^{f}+d_{I_{n}I_{p}}\sum_{j}R_{ij}^{I_{n}I_{p}}R_{fj}^{I_{n}I_{p}}\nonumber
\end{equation}
\begin{equation}
+N_{I_{p}}^{i}\sum_{j}R_{fj}^{I_{n}I_{p}}M_{j}^{-}+N_{I_{n}}^{f}\sum_{j}R_{ij}^{I_{n}I_{p}}M_{j}^{+}].
\label{Eqt. 20}
\end{equation}
\begin{equation}
M_{\beta^{+}}=\langle\psi_{I_{p}K_{p}}^{f}|\beta_{\mu}^{+}|\psi_{I_{n}K_{n}}^{i}\rangle\nonumber
\end{equation}
\begin{equation}
=-[\bar{d}_{I_{n}I_{p}}N_{I_{n}}^{i}N_{I_{p}}^{f}+d_{I_{n}I_{p}}\sum_{j}R_{ij}^{I_{n}I_{p}}R_{fj}^{I_{n}I_{p}}\nonumber
\end{equation}
\begin{equation}
+N_{I_{n}}^{i}\sum_{j}R_{fj}^{I_{n}I_{p}}M_{j}^{+}+N_{I_{p}}^{f}\sum_{j}R_{ij}^{I_{n}I_{p}}M_{j}^{-}],
\label{Eqt. 21}
\end{equation}
where $\mu=K_{f}-K_{i}$.$d_{np}$ and $\bar{d}_{np}$ are also reduced
matrix elements and they are given by
\begin{eqnarray}
d_{np}=\frac{1}{\sqrt{3}}u_{j_{p}}u_{j_{n}}<j_{n}\parallel\sigma\parallel
j_{p}>\nonumber
\end{eqnarray}
\begin{eqnarray}
\bar{d}_{np}=\frac{1}{\sqrt{3}}v_{j_{n}}v_{j_{p}}<j_{n}\parallel\sigma\parallel
j_{p}>
\end{eqnarray}
The reduced transition probability for the
$I_{i}K_{i}\longrightarrow I_{f}K_{f}$ transitions in the laboratory
frame is expressed by
\begin{equation}
B_{GT}^{\pm}(I_{i}K_{i}\rightarrow I_{f}K_{f})=\nonumber
\end{equation}
\begin{equation}
\frac{g_{A}^{2}}{4\pi}(I_{i}K_{i}1K_{f}-K_{i}/I_{f}K_{f})^{2}|M_{\beta^{\pm}}|^{2},
\label{Eqt. 22}
\end{equation}

The formalism used in PM is also used in Schematic Model (SM) with
one major difference. The effective interaction term ($h_{0}$) is
not added to the total Hamiltonian  in the SM (for further details,
see Refs. \cite{Bab05, Bab05a, Nec10, sel04, Sal03}).

\subsection{The pn-QRPA Method}
The Hamiltonian of the pn-QRPA model is given by
\begin{equation}
H^{QRPA} = H^{sp} + V^{pair} + V ^{ph}_{GT} + V^{pp}_{GT},
\label{Eqt. 23}
\end{equation}
and it is diagonalized as outlined below. Single particle energies
and wave functions are calculated in the Nilsson model which takes
into account nuclear deformation (for our Model (C)).  Pairing is
treated in the BCS approximation. Details of these two steps can be
seen from Ref. \cite{Hir93} and they are not reproduced here to save
space.

In the pn-QRPA formalism, GT transitions are expressed in terms of
phonon creation and one defines the QRPA phonons as
\begin{equation}
A^{+}_{\omega}(\mu) =
\sum_{pn}(X^{pn}_{\omega}(\mu)a^{+}_{p}a^{+}_{\overline{n}} -
Y^{pn}_{\omega}(\mu)a_{n}a_{\overline{p}}). \label{Eqt. 24}
\end{equation}
The sum in Eq.~(\ref{Eqt. 24}) runs over all proton-neutron pairs
with $\mu = m_{p} - m_{n}= -1, 0, 1,$ where $m_{p/n}$ denotes the
third component of the angular momentum. The ground state of the
theory is defined as the vacuum with respect to the QRPA phonons,
$A_{\omega}(\mu)|QRPA \rangle = 0$. The forward and backward-going
amplitudes $X$ and $Y$ are eigenfunctions of the RPA matrix equation
\begin{equation}
\left[
\begin{array}{cc}
A& B\\
-B & -A
\end{array}\right]
 \left[
\begin{array}{c}
X\\
Y
\end{array}\right]=\omega\left[
\begin{array}{c}
X\\
Y
\end{array}\right],
\label{Eqt. 25}
\end{equation}
where $\omega$ are energy eigenvalues. Again we refer to
\cite{Hir93} and references therein for solution of the RPA
equation~(\ref{Eqt. 25}).

The proton - neutron residual interaction occurs through two
channels: pp and ph channels. Both the interaction terms can be
given a separable form. The ph force is given by
\begin{equation}
V^{ph}_{GT} = 2\chi\sum_{\mu}(-1)^{\mu}Y_{\mu}Y^{+}_{-\mu},\nonumber
\end{equation}
with
\begin{equation}
Y_{\mu} = \sum_{j_{n}j_{p}} \langle
j_{p}m_{p}|t_{-}\sigma_{\mu}|j_{n}m_{n} \rangle
c^{+}_{j_{p}}m_{p}c_{j_{n}m_{n}}, \label{Eqt. 26}
\end{equation}
whereas the pp interaction given by the separable force
\begin{equation}
V^{pp}_{GT} =
2\kappa\sum_{\mu}(-1)^{\mu}P_{\mu}P^{+}_{-\mu},\nonumber
\end{equation}
 with
\begin{equation}
P_{\mu}^{+} = \sum_{j_{n}j_{p}} \langle
j_{n}m_{n}|(t_{-}\sigma_{\mu})^{+}|j_{p}m_{p} \rangle \nonumber
\end{equation}
\begin{equation}
\times
(-1)^{l_{n}+j_{n}-m_{n}}C^{+}_{j_{p}m_{p}}C^{+}_{j_{n}-m_{n}},
\label{Eqt. 27}
\end{equation}
is taken into account (in Models (B) and (C)). The interaction
constants $\chi$ and $\kappa$ in units of MeV are both taken to be
positive. The different sings of $V^{pp}$ and $V^{ph}$ reflect a
well-known feature of the nucleon-nucleon interaction: the ph force
is repulsive while the pp force is attractive. For further details
see Ref. \cite{Hir93}. The reduced transition probabilities for GT
transitions from the QRPA ground state to one-phonon states in the
daughter nucleus are obtained as
\begin{equation}
B_{GT}^{\pm}(\omega) = |\langle \omega, \mu
\|t_{\pm}\sigma_{\mu}\|QRPA\rangle|^{2}. \label{Eqt. 28}
\end{equation}
For odd-A nuclei, there exist two different types of transitions:
(a) phonon transitions with the odd particle acting only as a
spectator and (b) transitions of the odd particle itself. For case
(b) phonon correlations are introduced to one-quasiparticle states
in first-order perturbation. For further details, we refer to
\cite{Hir93}.

In order to improve the reliability of calculated results in pnQRPA
(C) and SM (c) models, experimentally adopted value of the
deformation parameter for $^{42,44,46,48,50}$Ti, extracted by
relating the measured energy of the first $2^{+}$ excited state with
the quadrupole deformation, was taken from Raman et al.
\cite{Ram87}. For all other cases, where measurement has not been so
far done, the deformation of the nucleus was calculated using
\begin{equation}
\delta = \frac{125(Q_{2})}{1.44 (Z) (A)^{2/3}},
\end{equation}
where $Z$ and $A$ are the atomic and mass numbers, respectively and
$Q_{2}$ is the electric quadrupole moment taken from Ref.
\cite{Moe81}. Q-values were taken from the recent mass compilation
of Audi et al. \cite{Aud03}.

\section{GT$_{\pm}$ Strength Distributions}
In a sense both $\beta$-decay and capture rates are very sensitive
to the location of the GT$_{+}$ centroid. An $(n,p)$ experiment on a
nucleus $(Z,A)$ shows the place where in $(Z-1,A)$ the GT$_{+}$
centroid corresponding to the ground state of $(Z,A)$ resides. The
$\beta$-decay and electron capture rates are exponentially sensitive
to the location of GT$_{+}$ resonance while the total GT strength
affect the stellar rates in a more or less linear fashion
\cite{Auf96}. Each excited state of $(Z,A)$ has its own GT$_{+}$
centroid in $(Z-1,A)$ and all of these resonances must be included
in the stellar rates. We do not have the ability to measure these
resonances. Similar is the case in the $\beta^{-}$ direction. Here,
every excited state of $(Z,A)$ also has its own GT$_{-}$ centroid in
$(Z+1,A)$ and again all the contributions should be included in a
reliable estimate of stellar $\beta^{-}$-decay rates. Turning to
theory, we see that the pioneer calculation done by Fuller and
collaborators \cite{Ful82} (referred to as FFN throughout this text)
had to revert to approximations in the form of Brink's hypothesis
and "back resonances" to include all resonances in their
calculation. Brink's hypothesis states that GT strength distribution
on excited states is \textit{identical} to that from ground state,
shifted \textit{only} by the excitation energy of the state. GT back
resonances are the states reached by the strong GT transitions in
the inverse process (electron capture) built on ground and excited
states. Even the microscopic large-scale shell model calculations
\cite{Lan00} had to use the Brink assumption to include all states
and resonances. On the other hand, the pn-QRPA model is the only
model that provides a microscopic way of calculating the GT$_{\pm}$
centroid and the total GT$_{\pm}$ strength for \textit{all} parent
excited states and it can lead to a fairly reliable estimate of the
total stellar rates. The PM and SM have so far not been used to
calculate excited state GT strength functions.

In this section, our calculation results for GT strength
distribution function by using different models (pn-QRPA, PM and SM)
are given. As discussed previously, we sub-divide the pn-QRPA and SM
into three categories (A), (B) and (C). Model (A) is the most basic
model in which only the interaction in ph channel is considered.
Model (B) is to check the difference in calculations when one also
incorporates pp force. Model (C) highlights the dependence of QRPA
calculations on nuclear deformations. PM is classified only into two
categories namely (A) and (B). The GT strength extracted from (p,n)
spectra is about 40$\%$ lower than the Ikeda sum rule \cite{Gaa84}.
Two possible mechanisms behind quenching of GT strength are pure
nucleonic mechanism and $\Delta$ mechanism. For further details we
refer to \cite{Gro90}. El-Kateb and collaborators \cite{Kat94}
employed a much smaller quenching factor of 0.23 for $^{55}$Mn, 0.31
for $^{56}$Fe and $^{58}$Ni for strength below 10 MeV excitation to
compare shell model calculation with the measured data. Our
calculated GT strengths are all quenched within the pn-QRPA
formalism by universal factor of
 $f_{q}^{2}$ = (0.6)$^{2}$ which is used for fp shell nuclei ( also employed in Ref. \cite{Nab11}).
The re-normalized Ikeda sum rule in pn-QRPA is given by
\begin{equation}
ISR_{renorm}=S^{(-)}-S^{(+)}\cong 3f_{q}^{2}(N-Z). \label{Eqt. 34}
\end{equation}

Rather than presenting the detailed GT strength distributions for
the twenty-one isotopes of titanium using the eight different models
mentioned above, the key statistics of GT strength distribution
(total strength, centroid and width) are shown in Tables 1 to 4. It
is noted that the re-normalized Ikeda sum rule is fulfilled by
pn-QRPA models (deviations are within a few percent and are
attributed to non-nucleonic effects). The Ikeda Sum Rule in SM and
PM models is given by Eq. 13. Total strength calculations are
performed up to 20 MeV in pn-QRPA whereas in PM and SM models they
are calculated up to 40 MeV.

Table~\ref{tbl-1} displays the calculated total GT strengths,
centroids and widths for isotopes of titanium ($^{40-44}$Ti) in both
$\beta$-decay and electron capture directions. Results are shown for
all eight models (which are also explained in the footnote of
Table~\ref{tbl-1}). Centroids and widths are given in units of MeV
and B(GT) strengths are given in units such that B(GT) = 3 for
neutron decay.

As seen from Table~\ref{tbl-1}, the calculated values for $^{40}$Ti
show that PM and SM models give much bigger values for total GT
strength in electron capture direction. The centroids  are also
placed at higher excitation energies in daughter nuclei for the PM
and SM models. These two models also generally calculate bigger
widths than the corresponding results of pn-QRPA model.  The effect
of the inclusion of the pp interaction in pinning down the centroid
values is more pronounced in PM model (they approximately decrease
to half their original values). However, pp force does not show any
similar change in other calculated quantities by remaining two
models. The pn-QRPA model calculates the lowest centroids and
widths.

For the case of odd-A nucleus $^{41}$Ti, one notes that for electron
capture direction, the pn-QRPA (C) calculates bigger total strength
as compared to pn-QRPA (A) and (B). In the $\beta$-decay direction,
the three pn-QRPA models show close results. The SM (C) and PM (B)
models bring substantial improvement over SM (A), SM (B) and PM (B)
models leading to much lower values of centroids and widths in
$\beta$-decay direction and bigger total GT strength values. SM (C)
calculates lowest centroid and biggest strengths for $^{41}$Ti.

The B(GT) values, centroids and widths for $^{42}$Ti calculated by
the pn-QRPA model decrease when the pp interaction for spherical
case is taken into account (compare versions (A) and (B)), albeit
not much. Deformation substantially changes the pn-QRPA results for
total strength and centroid. PM (B) brings down the calculated
centroid values roughly by a factor of three and at the same time
increases the total strength and width values in both directions.
The pp force in PM method pins down the centroid values but results
in no significant changes in the calculated values of total strength
and width. The role of pp force in SM model for spherical case
increases the total strengths, centroids and widths. However, when
the pp force is considered together with deformation, the centroids
and widths are roughly halved and at the same time there is an
increase in total strength values. The pp force in all models
assists in shifting the centroid to lower excitation energies in
daughter.

No appreciable difference is seen among the three pn-QRPA models for
the case of $^{43}$Ti specially in the $\beta$-decay direction. On
the other hand, SM (C) not only substantially changes the results of
SM (A) and SM (B) but also calculates lower widths and centroids as
compared to pn-QRPA (C). Moreover, SM (C) is also able to calculate
bigger total strength, especially in $\beta$-decay direction. PM (A)
and PM (B) give smaller B(GT) values than pn-QRPA and SM models. The
effect of the incorporated deformation is significant in the case of
SM.

The calculated results for $^{44}$Ti have been presented as a last
entry in Table 1. The biggest and lowest values for total strengths
have been obtained by the SM (C) and PM (A) models, respectively. PM
model values calculate very small GT strength. All three models do
not exhibit any substantial change in calculated quantities in the
given versions. Deformation tends to increase total strength,
centroid and width values in SM model. SM and pn-QRPA models have
around three times larger B(GT) values than the PM models. For the
case of centroids and widths, the results of PM models are
approximately three times bigger than the corresponding ones in SM
and pn-QRPA models.

The key statistics for calculated GT transitions in $^{45-50}$Ti are
presented in Table 2. For $^{45,47,49}$Ti there is no much
appreciable change in the results of all quantities calculated by
pn-QRPA and PM models for $\beta^{-}$ and $\beta^{+}$ decay. For the
case of $^{45}$Ti, PM (A), PM (B) and SM (B) calculate small total
strength in both directions. The effect of the pp force does not
change significantly the results in PM model. Deformation provides a
drastic decrease in widths of the calculated GT distributions for SM
model. The centroid values in PM model are appreciably low. On the
other hand, the calculated width values in PM models for $\beta^{-}$
and $\beta^{+}$ decays are bigger than the corresponding results in
SM and pn-QRPA models.  For $^{47}$Ti, SM (C) makes a significant
change in the centroid, width and total strength values.  The effect
of the incorporated deformation in the three models is most
pronounced in SM and least in pn-QRPA. For $^{49}$Ti, results are
similar as in the case of $^{45,47}$Ti. When deformation is taken
into account the B(GT) values for $\beta$-decay in pn-QRPA and SM
models increase by a factor of 2-3. PM models gives a very small
total strength for $\beta^{+}$ decay. For the case of even-even
isotopes ($^{46-50}$Ti), it is noted that no pronounced difference
occurs among the results of pn-QRPA (A) and pn-QRPA (B) versions.
One notes that pn-QRPA (B) version gives lower B(GT) and width
values for both $\beta^{+}$ and $\beta^{-}$ decay in $^{46,48}$Ti.
Vanishing total strength in $\beta^{+}$ direction is calculated for
$^{48}$Ti in PM models. In $^{50}$Ti isotope, SM (C) model gives the
highest centroid and width values amongst all other models.

In Table~3, the calculated total GT strengths, centroids and widths
for $^{51-56}$Ti are shown for the three models. For
$^{51,53,55}$Ti, SM (C) calculates biggest total strength along
$\beta$-decay direction. For the odd-A cases, SM (C) calculates the
biggest total strength. The SM models tend to calculate largest
centroid values along $\beta$-decay direction. The pn-QRPA models
calculate the lowest centroid in $\beta$-decay direction for all Ti
isotopes in Table~3.

Table~4 finally displays the calculated total strengths, widths and
centroids for $^{57-60}$Ti. Isotopes of titanium gets progressively
neutron-rich as one proceeds from Table 1 to Table 4. In Table~4,
pn-QRPA (C) model calculates low centroids and reasonable GT
strengths (satisfying renormalized Ikeda sum rule Eq. 29). PM model
places centroids for $\beta^{-}$-decay centroids in $^{58,60}$Ti at
much lower energies than the SM model. It can be seen that the total
strength values in PM (A) and PM (B) models are very closer to the
corresponding ones in SM (A) and SM (B) for $^{58,60}$Ti. In
$^{57,59}$Ti, the calculated GT strength values increases
substantially in SM (C) in accordance with the Ikeda sum rule.

Tables~\ref{tbl-1} to \ref{tbl-4} show that the pn-QRPA models
follow systematic trend in the calculation of GT strength function
for both even-even and odd-A nuclei. The values of pn-QRPA
calculated total strength decreases (increases) systematically,
separately for even-even and odd-A nuclei, along the electron
capture ($\beta$-decay) direction. This trend is valid only for
even-even nuclei in PM and SM models. The pn-QRPA (C) is the best
model for the calculation of GT strength distribution amongst all
eight models presented in this work for even-even and odd-A nuclei.
The pn-QRPA (C) model calculates reasonable total strength in both
directions for all cases (see comparison with measured data below).
Moreover, the model calculates lower centroid energies in daughter
which translates into bigger weak interaction rates and can bear
consequences for astrophysical applications. Further, it is only the
pn-QRPA (C) model that fulfills the re-normalized Ikeda sum rule
Eqt.~(\ref{Eqt. 34}). Only in the case of $^{55}$Ti is the
re-normalized sum rule satisfied to only 94$\%$.  The pn-QRPA (A) is
not able to satisfy the sum rule for few odd-A cases whereas version
(B) fails for few mixed cases. However, PM (B) model also shows good
results of GT strength distributions for even-even nuclei. The PM
(B) and SM (C) tend to fulfill the Ikeda sum rule Eqt.~(\ref{Eqt.
13}) for even-even nuclei.  The PM (B) and SM (C) are the better
model in its genre and shows overall better results in their class.
The results support the argument that the QRPA models perform best
when performed both in \textit{pp} and \textit{ph} channels, taking
nuclear deformation into consideration.

So far we have only shown the mutual comparison of the calculated
total GT strength distributions amongst the eight theoretical
models. It would be interesting to see the comparison of the results
for these models with the measured data where available. For this
reason, we searched the literature and were able to find at least
five cases of reported measured GT strength distributions of
titanium isotopes ($^{40,41,46,47,48}$Ti). Next, we compare the
measured data with the the results of three preferred models, namely
pn-QRPA (C), SM (C) and PM (B). Moreover, we also compare our
results for these cases with other theoretical calculations.

\begin{figure}
\includegraphics [height=5in,width=2.5in]{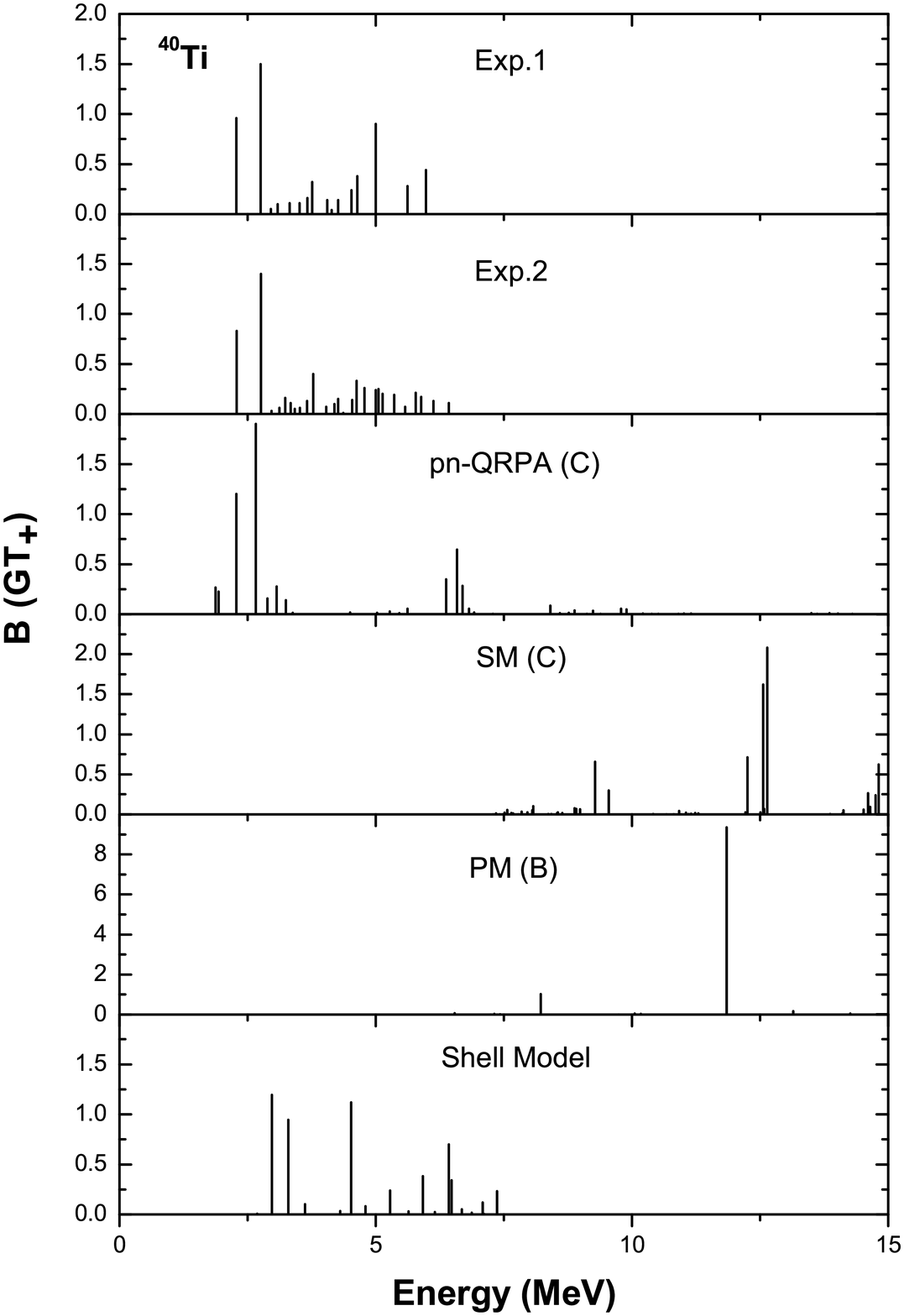}
\caption{Comparison of Gamow-Teller (GT$_{+}$) strength
distributions for $^{40}$Ti $\beta$ decay. Measured data Exp. 1 and
Exp. 2 were taken from \cite{Tri97} and \cite{Liu98}, respectively.
Shell Model  results were taken from Ref. \cite{Orm95}. }
\label{fig1}
\end{figure}

Fig.~\ref{fig1} shows the GT$_{+}$ strength distribution for
$^{40}$Ti. Liu and collaborators \cite{Liu98} studied the $\beta$
decay of $^{40}$Ti and $^{41}$Ti and its subsequent implication for
detection of solar neutrinos. Trinder et al. also made a study of
the $\beta$-decay measurement of $^{40}$Ti \cite{Tri97} at GANIL
using the LISE3 spectrometer. For details of experiment, we refer to
\cite{Tri97}. The results of these two $\beta$-decay experiments are
shown in top two panels of Fig.~\ref{fig1}. In the top panel, Exp. 1
shows the measured data of Ref. \cite{Tri97} while Exp. 2
corresponds to that of Ref. \cite{Liu98}. The third, fourth and
fifth panels show our calculated results of pn-QRPA (C), SM (C) and
PM (B) models, respectively. We also show the (0+2)$\hbar\omega$
shell model calculation of GT$_{+}$ strength distribution for
$^{40}$Ti in bottom panel performed by Ormand and collaborators
\cite{Orm95}. In order to bring their calculated GT strength in
compliance with the measured GT data, the authors re-normalized the
free nucleon GT operator by the factor 0.775 \cite{Orm95}. The
horizontal axis in Fig.~\ref{fig1} shows the energy scale in
daughter $^{40}$Sc in units of MeV. It can be seen from
Fig.~\ref{fig1} that the pn-QRPA (C) model reproduces well the
low-lying measured GT strengths of Refs. \cite{Liu98, Tri97} and
also predicts some GT transitions above 8 MeV in daughter which are
not reported by measurements. Shell Model calculation \cite{Orm95}
is in good agreement with the measured data. The pn-QRPA (C)
calculated total strength, up to 15 MeV in daughter, is 6.00 and is
in very good agreement with the measured strength of 5.86 (5.87) by
Ref. \cite{Liu98} (Ref. \cite{Tri97}). Shell Model calculated a
total strength of 5.62. The SM (C) and PM (B) models give a total
strength value of 7.62 and 10.80, up to 15 MeV, and they are
considerably bigger than the experimental values. The shell model
data is not well fragmented as compared to pn-QRPA (C) data. This is
possibly due to the neglect of higher-order correlations in the
shell model. The pn-QRPA (C) model placed the GT centroid at 3.91
MeV in daughter and it is also in very good agreement with the
measured data of 3.87 MeV by Ref. \cite{Liu98} and 3.78 MeV by Ref.
\cite{Tri97}. Shell model placed the centroid at 4.67 MeV whereas
the PM(B) and SM (C) models placed the centroid at a too high
excitation energy of 11.47 MeV and 12.17 MeV, respectively.

\begin{figure}
\includegraphics [height=5in,width=2.5in]{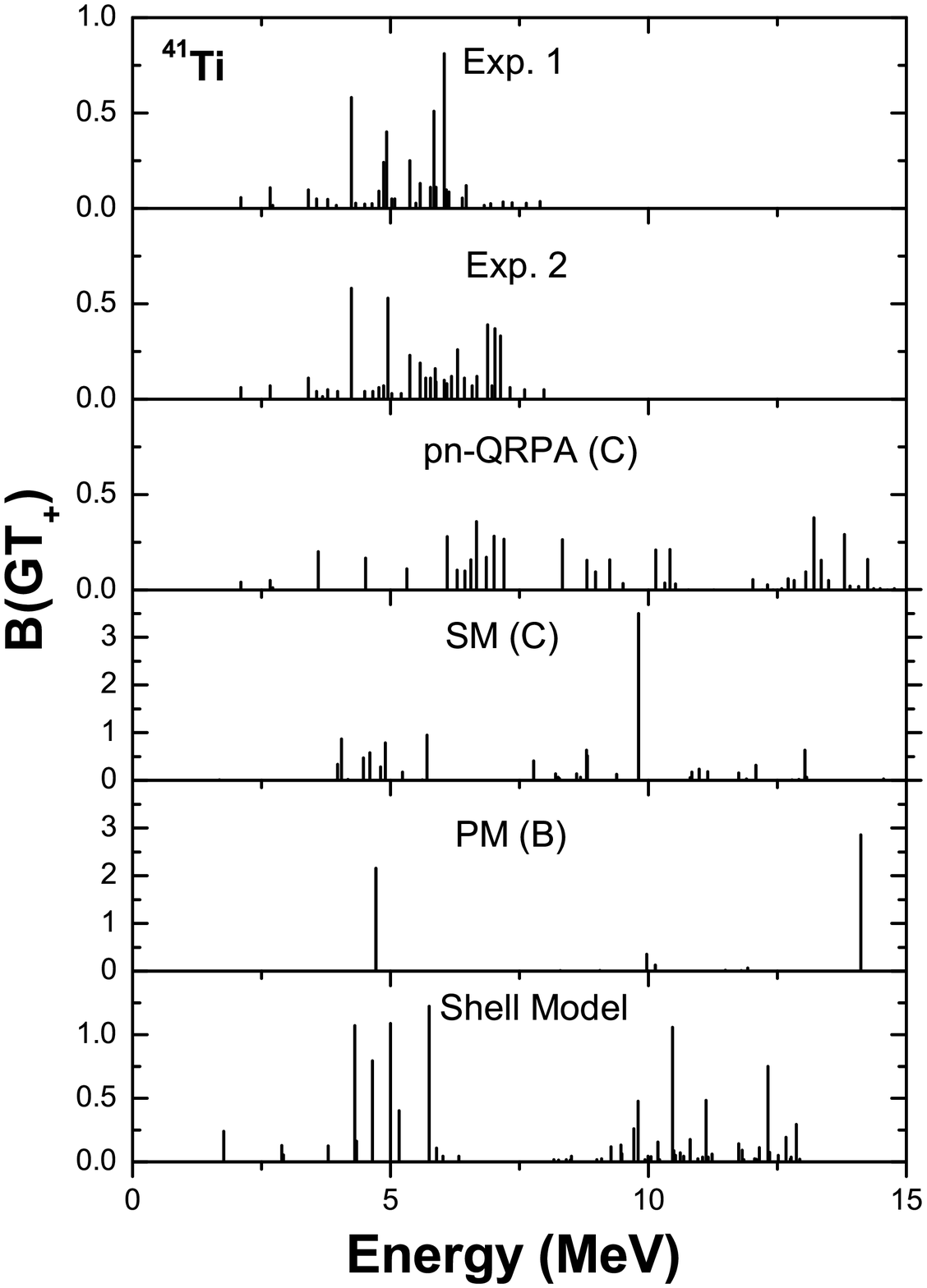}
\caption{Comparison of Gamow-Teller (GT$_{+}$) strength
distributions for $^{41}$Ti $\beta$ decay. Measured data Exp. 1 and
Exp. 2 were taken from \cite{Honk97} and \cite{Liu98}, respectively.
Shell Model results were taken from Ref. \cite{Honk97}.}
\label{fig2}
\end{figure}

Honkanen et al. \cite{Honk97} performed an improved high-resolution
study of the $\beta$-decay of $^{41}$Ti produced in the
$^{40}$Ca($^{3}$He,2n) reaction at 40 MeV at the IGISOL facility
\cite{Honk97}. In addition, the authors also performed a shell model
calculation of the GT strength distribution of $^{41}$Ti in the
\textit{sdfp} space. For details of the experiment and theoretical
shell model calculation, we refer to \cite{Honk97}. We show the
measured data by Honkanen and collaborators \cite{Honk97} as Exp. 1
in the top panel of Fig.~\ref{fig2}. The measured GT strength
distribution of $^{41}$Ti by Liu et al. \cite{Liu98} is shown in
second panel as Exp. 2.  The next three panels show our calculated
results for pn-QRPA (C), SM (C) and PM (B) models, respectively.  In
the bottom panel, we show the shell model calculation of Ref.
\cite{Honk97}. The measured data is well fragmented up to 8 MeV in
daughter. The theoretical models calculate a well fragmented data,
akin to measured data. All theoretical models do calculate
high-lying GT transitions not reported by experiments. The Shell
Model data calculates much bigger total strength of 10.92 which is
to be compared with measured strength of 4.83 by Liu et al. and 4.34
by Honkanen and collaborators. The pn-QRPA (C) model calculates a
total strength of 4.89 in excellent agreement with the measured
value of 4.83 \cite{Liu98}. The total strength calculated by the SM
(C) is 4.36 in excellent agreement with the measured value of 4.34
\cite{Honk97} whereas the PM (B) model finds a total strength of
7.76. The centroid in the SM (C) model is placed at too high
excitation energy of 8.13 MeV in $^{41}$Sc. The pn-QRPA (C)
calculated the centroid value of 8.96 MeV which can be compared with
the shell model results of 7.86 MeV . In comparison, the centroids
of measured data by Refs. \cite{Honk97} and \cite{Liu98} are at 5.27
MeV and 5.67 MeV, respectively. The PM (B) model gives a centroid
value of 8.61 MeV. It is to be noted that whereas the measured data
is available only up to an excitation energy of 8 MeV, the
theoretical data are given up to an excitation energy of 15 MeV in
daughter. If one cuts the theoretical data also till 8 MeV then the
calculated centroids can come in reasonable agreement with the
measured centroids. Fig.~\ref{fig2} shows that a quenching of GT
strength calculated by shell model is in order. Experimentalists are
urged to search for high-lying GT transitions (up to 15 MeV) in
$^{41}$Sc.

\begin{figure}
\includegraphics [height=5in,width=2.5in]{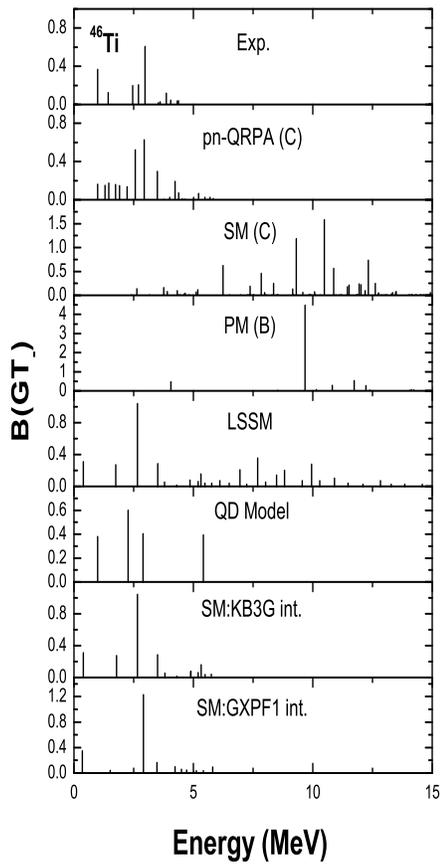}
\caption{Comparison of Gamow-Teller (GT$_{-}$) strength
distributions for $^{46}$Ti $\beta$-decay. Measured data Exp. was
taken from \cite{Ada06}. Large Scale Shell Model (LSSM) results were
taken from Ref. \cite{Pet07}, whereas Shell Model:KB3G int. and
Shell Model:GXPF1 int. were taken from Ref. \cite{Pov01} and
\cite{Hon04}, respectively. Quasi Deutron (QD) model data was taken
from Ref. \cite{Lis01}. } \label{fig3}
\end{figure}

For the case of $^{46}$Ti, there are quite a few theoretical GT
calculations available in the $\beta$-decay direction. The results
are shown in Fig.~\ref{fig3} which comprises of eight panels. The
measured data was taken from the recent $\beta$-decay measurement of
$^{46}$Ti by Adachi and collaborators \cite{Ada06} and it is shown
in the top panel of Fig.~\ref{fig3}. The authors performed a
high-resolution ($^{3}$He,t) experiment on $^{46}$Ti at 0$^{0}$ and
at an intermediate incident energy of 140 MeV/nucleon for the study
of precise measurement of GT transitions in $^{46}$V. A very good
energy resolution of $\Delta E \le $ 50 keV was realized in the
experiment. For further details of the experiment, Ref. \cite{Ada06}
can be seen. Besides our pn-QRPA (C), SM (C) and PM (B) models
(shown in second, third and fourth panel, respectively), we also
show the results of GT strength distributions from four other
theoretical calculations. The large scale shell model (LSSM)
calculation of Petermann et al. \cite{Pet07} is shown in the fifth
panel. Petermann and collaborators used the KB3G interaction
\cite{Pov01} and employed Lanczos method with 100 iterations
ensuring convergence in their calculation for 1-2 MeV excitation
energies. The sixth panel shows the quasideutron (QD) model
calculation with a deformed core (i.e. rotor $+$ quasideutron model)
performed by Lisetskiy et al \cite{Lis01}. The last two panels show
shell model calculations using KB3G \cite{Pov01} and GXPF1
\cite{Hon04} interactions, respectively. All shell model data as
well as the QD model used a quenching factor of (0.74)$^{2}$ in
their calculations. It can be seen from Fig.~\ref{fig3} that the
pn-QRPA (C) data is well fragmented and it is in good agreement with
the $\beta$-decay measurement performed by Adachi et al. The pn-QRPA
(C) model also calculates its strongest GT transition around 3 MeV
akin to measured data. The SM (C) model calculates the fragmented GT
strength distribution and high-lying transitions in the range of
6-12 MeV. The largest GT strength which is 4-8 times larger than
other calculations and experimental result is obtained by PM (B)
model. In this model, only one GT peak is seen around 9 MeV. The
shell model results do not produce enough fragmentation of GT
strength, specially at low excitation energies (between 1 and 3
MeV). The QD model calculates only four transitions. In QD model,
the low-lying states in $^{46}$V are described by an
angular-momentum-coupled proton-neutron pair (quasideutron) made of
the valence odd proton and neutron occupying Nilsson orbits coupled
to the rotating $^{44}$Ti core. Table~\ref{tbl-5} presents the total
GT strength as well as the calculated centroid for all GT
distribution functions shown in Fig.~\ref{fig3}.  It can be seen
from Table~\ref{tbl-5} that the QD model best reproduces the
measured total GT strength up to 5.4 MeV in $^{46}$V. Once again it
is to be noted that LSSM, pn-QRPA (C), PM (B) and SM (C) have a high
cutoff in daughter excitation energy and reducing this cutoff can
lead to much better comparison with measured data. At the same time,
experimentalists can be directed to perform their measurement up to
around 15 MeV in daughter to look for more high-lying GT strength in
$^{46}$V.

\begin{figure}
\includegraphics [height=5in,width=2.5in]{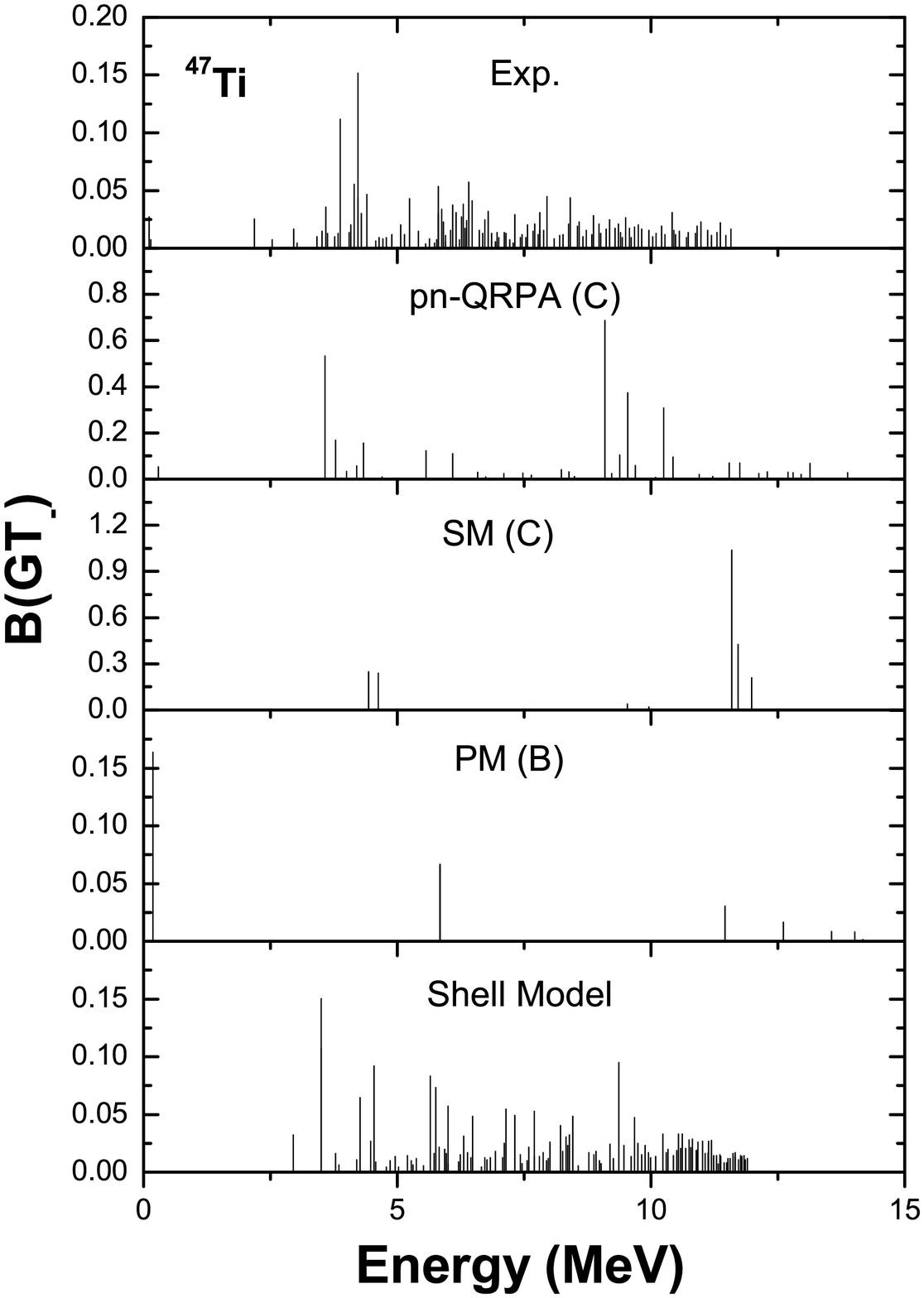}
\caption{Comparison of Gamow-Teller (GT$_{-}$) strength
distributions for $^{47}$Ti $\beta$-decay. Measured data Exp. was
taken from \cite{Gan13}. Shell Model results were taken from Ref.
\cite{Hon04}. } \label{fig4}
\end{figure}

A high resolution ($^{3}$He,t) experiment for $^{47}$Ti was
performed at the Research Center for Nuclear Physics, Japan at an
intermediate incident energy of 140 MeV/nucleon and a very fine
energy resolution of 20 keV. The measurement was reported recently
\cite{Gan13} up to an excitation energy of 12.5 MeV in $^{47}$V. The
authors were suggestive that high-lying GT strength beyond 12.5 MeV
might also exist well . The results of the measurement are shown in
the top panel of Fig.~\ref{fig4}. Once again we depict the
calculated results of pn-QRPA (C), SM (C) and PM (B) in the second,
third and fourth panel of Fig.~\ref{fig4}, respectively. The bottom
panel finally shows the shell model calculation  using the GXPF1
interaction \cite{Hon04} including a quenching factor of
(0.74)$^{2}$. The shell model and experimental results are generally
in agreement. However, above 10 MeV, the shell model cumulative sum
is larger than the experimental one. The pn-QRPA (C) data is also
fragmented like the shell model and experimental data and it is in
excellent agreement with the measured data. The SM (C) and PM (B) do
not perform well for this odd-A isotope of titanium. The PM (B)
calculates a peak value of 0.16 at a relative low excitation energy
of 0.2 MeV whereas the SM (C) calculates the biggest transition of
1.04 magnitude at 11.6 MeV. The pn-QRPA (C) calculated a total
strength of 3.58 in complete agreement with the measured value of
3.60. The corresponding value calculated within the shell model is
2.76. The SM (C) calculates yet smaller value of total strength of
2.47 whereas the PM (B) bags a paltry sum of 0.36. Regarding
centroid placement, one notes that pn-QRPA (C) calculates the
centroid at 7.62 MeV which is again in excellent agreement with the
measured centroid of 7.5 MeV. LSSM fixes the centroid at 7.59 MeV.
The SM (C) places the centroid at a higher value of 9.48 MeV and PM
(B) places it at 4.45 MeV.

\begin{figure}
\includegraphics [height=5in,width=2.5in]{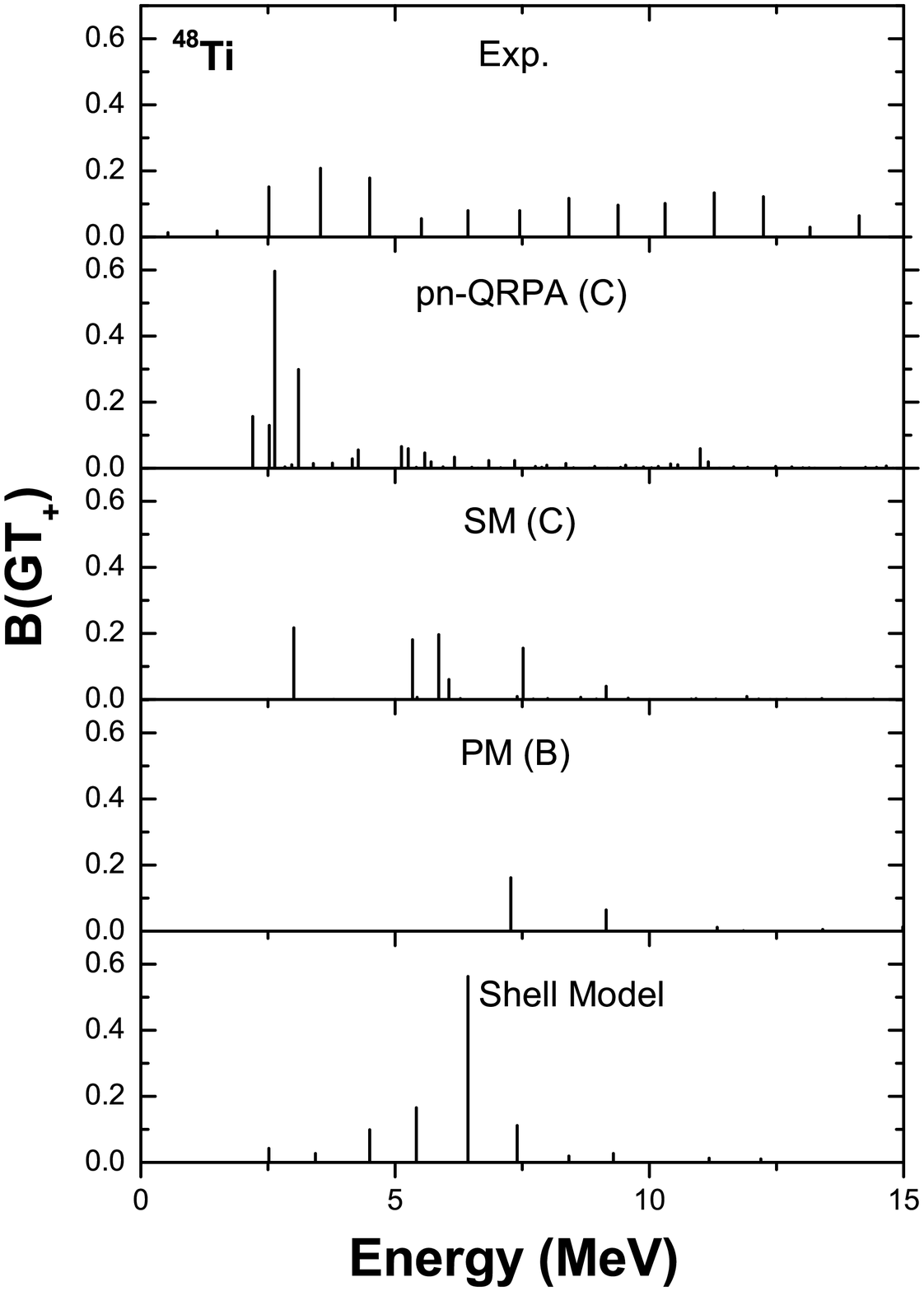}
\caption{Comparison of Gamow-Teller (GT$_{+}$) strength
distributions for $^{48}$Ti $\beta$-decay. Measured data Exp. was
taken from \cite{Alf90} whereas Shell Model (SM) results were taken
from Ref. \cite{Bro85}.} \label{fig5}
\end{figure}

We finally present the GT strength distributions of $^{48}$Ti (in
the electron capture direction) in Fig.~\ref{fig5}. Alford and
collaborators \cite{Alf90} studied the $^{48}$Ti(n,p) reaction at an
energy of 200 MeV and they were able to obtain GT strength
distribution of $^{48}$Ti up to a comparatively much higher energy
value of around 12 MeV in daughter nucleus, $^{48}$Sc. Further
details of the performed experiment can be seen from Ref.
\cite{Alf90}. The top panel of Fig.~\ref{fig5} shows the measured GT
distribution obtained from the (n,p) reaction experiment. This is
followed by our model calculations of pn-QRPA (C), SM (C) and  PM
(B). The bottom panel presents the shell model calculation by Brown
\cite{Bro85} in a model space
$(f_{7/2})^{8-n}(f_{5/2}p_{3/2}p_{1/2})^{n}$ with $n$ = 0, 1, and 2.
The shell model data has been quenched by a factor 0.6. It can be
seen from Fig.~\ref{fig5} that the measured GT strength is well
fragmented and extend to high excitation energies in $^{48}$Sc. The
pn-QRPA (C) calculation is also well fragmented. The first measured
peak at 2.52 MeV is well reproduced by the pn-QRPA (C) model but it
also calculates a strong peak of 0.6 at 2.64 MeV  in $^{48}$Sc. The
SM (C) data is also fragmented but is off much lower strength. The
shell model calculates the strongest peak of 0.6 at 6.44 MeV. The
total measured GT strength of 1.44 is to be compared with the
theoretical values of 1.78, 0.90, 0.26 and 1.08 by pn-QRPA (C), SM
(C), PM (B) and shell model, respectively.  The centroid of the
measured GT distribution resides at 7.31 MeV in daughter. The
pn-QRPA (C) model calculates a much lower centroid at 4.12 MeV. The
SM (C) locates the centroid at 5.72 MeV whereas the shell model
places it at 6.19 MeV in daughter $^{48}$Sc. The highest centroid
value has been obtained by PM (B) model which is 8.47 MeV. This
value is higher than the measured data although the corresponding
values for other theoretical calculations are lower than the
experimental results. The closest match with the measured centroid
value for the case of $^{48}$Ti is provided by shell model
calculation.

\section{Summary and conclusions}
Fermi and Gamow-Teller transitions are required for an accurate
calculation of $\beta$-decay and electron capture rates in
terrestrial and stellar environments. Reliable estimates of
$\beta$-decay half-lives (including many neutron-rich nuclei) are in
high demand in various nuclear physics (e.g. for the experimental
exploration of the nuclear landscape at existing and future
radioactive ion-beam facilities) and astrophysical problems (e.g.
for a better understanding of the supernova explosion mechanism and
heavy element nucleosynthesis).  It is the GT transitions which are
fragmented and are very challenging to calculate. The GT transitions
assume a nuclear model for its calculation whereas calculation of
Fermi transitions is straight forward (these are concentrated in a
single state known as isobaric analogue state). The Gamow-Teller
strength distribution for isotopes of titanium, $^{40-60}$Ti, were
calculated and analyzed by using three microscopic models (namely
the pn-QRPA, Pyatov Method and Schematic Method). The pn-QRPA and
Schematic model were further sub-divided into three classes in order
to highlight the role of particle-particle (pp) force and
deformation of the nucleus in the GT strength calculation within the
models. In Pyatov Method, only the effect of the pp force was
studied. The calculated GT strength functions were compared with the
corresponding experimental and other theoretical model calculations
wherever available. The isotopes of titanium chosen for this project
also included neutron-rich cases and have astrophysical
significance.

Our calculations strongly suggest that models with \textit{pp} force
and deformation of nucleus incorporated give better results for GT
strength distribution functions.

The calculations show that the inclusion of pp interaction and
deformation in pn-QRPA model tends to bring down the centroid
values. In PM and SM models, the pp force has a rather unpredictable
effect on centroid placement. Further, in general, for the case of
SM and PM models, the pp force does not show any sharp change in
width and GT strength calculations (see Tables~\ref{tbl-1} to
\ref{tbl-4}). Centroid placement in SM is generally at higher
excitation energies as compared to the PM even though the same
effective interaction constant values ($\chi_{ph}$ and $\chi_{pp}$)
are used in both models. Lower centroids  are attributed to the
inclusion of $h_{0}$ in PM models.

The calculated GT strength distribution functions were also compared
with measured GT distributions where available. Comparison with
other theoretical calculations were also sought in such cases. The
pn-QRPA (C) is best able to reproduce the measured strength and
centroid of $^{40}$Ti. The shell model calculated total strength is
also in good agreement with the measured strength of $^{40}$Ti. For
the case of $^{41}$Ti, both the pn-QRPA (C) and SM (C) models were
able to reproduce the measured total strength. All theoretical
models calculated high-lying GT transitions for $^{41}$Ti not
reported by measurements. The QD model best reproduces the measured
GT strength function in $^{46}$V. For $^{47}$Ti, the pn-QRPA (C)
best reproduces the measured total strength and centroid placement
(shell model calculated centroid is also in agreement with measured
data). For $^{48}$Ti, shell model is successful in reproducing the
measured GT strength function.

The pn-QRPA (C) model satisfied the re-normalized Ikeda sum rule to
within a few percent and calculated much bigger GT strengths and
lower centroids as compared to other QRPA models. The model also
followed a systematic trend in calculation of GT strength for
even-even and odd-A titanium isotopes. This trend is valid only for
even-even nuclei in PM and SM models. The pn-QRPA (C) model also
performed reasonably well in comparison to measured GT strength
distributions. SM (C) model also displayed encouraging results.

One expects significant progress in our understanding of supernova
explosions and heavy element nucleosynthesis to come from
next-generation radioactive ion-beam facilities (e.g. FAIR
(Germany), FRIB (USA) and FRIB (Japan)) when we would have access to
measured GT strength distribution of many more nuclei (including
unstable isotopes). Nonetheless, for astrophysical applications, one
needs microscopic and reliable calculation of GT strength
distributions for hundreds of iron-regime nuclei. We are in a
process of calculating GT strength functions for other key
\textit{fp}-shell nuclei (including many neutron-rich unstable
nuclei) in a microscopic fashion and hope to report our findings in
near future.

\acknowledgments S. Cakmak and T. Babacan would like to acknowledge
the support of research grant provided by BAP Project with number
2013-004. S. Cakmak would also like to acknowledge the kind
hospitality provided by the GIK Institute of Engineering Sciences
and Technology, Pakistan, where major portion of this project was
completed and manuscript written.

\onecolumn
\begin{table}
\small \caption{Total GT strengths, centroids and widths of
calculated GT strength distribution functions of titanium isotopes,
both in electron capture and $\beta$-decay directions, for various
QRPA models given in first column. For explanation of QRPA models
see footnote at the end of Table~\ref{tbl-1}.} \label{tbl-1}
\begin{tabular}{ccccccc}
 $\mathbf{^{40}}$\bf{Ti} & $\mathbf{\sum B(GT_{-})}$ & $\mathbf{\sum B(GT_{+})}$
& $\mathbf{\bar{E}_{-}}$ & $\mathbf{\bar{E}_{+}}$ & $\mathbf{Width_{-}}$ & $\mathbf{Width_{+}}$  \\
\tableline
pn-QRPA (A)\tablenotemark{a} & 0.78 & 4.95 & 6.51 & 4.16 & 3.08 & 2.43 \\
pn-QRPA (B)\tablenotemark{b} & 0.77 & 5.05 & 6.34 & 3.89 & 3.01 & 2.37 \\
pn-QRPA (C)\tablenotemark{c} & 1.77 & 6.04 & 6.41 & 3.97 & 3.02 & 2.37 \\
PM (A)\tablenotemark{d} & 0.65 & 13.20 & 16.49 & 10.92 & 7.05 & 2.29 \\
PM (B)\tablenotemark{e} & 0.65  & 11.81  & 8.94  & 5.69 & 7.71 & 2.97  \\
SM (A)\tablenotemark{f}& 0.43& 11.76 & 13.57 & 11.59 & 9.7 & 8.25 \\
SM (B)\tablenotemark{g}& 0.44 & 11.76 & 9.48 & 8.89 & 7.20 & 5.94\\
SM (C)\tablenotemark{h} & 0.28 & 9.01 & 16.48 & 12.62 & 4.80 & 3.01 \\
 $\mathbf{^{41}}$\bf{Ti} &  & &  &  &  &   \\
 \tableline
pn-QRPA (A) & 0.68 & 2.08 & 9.51 & 6.65 & 2.61 & 2.11\\
pn-QRPA (B) & 0.71 & 3.45 & 9.46 & 9.08 & 2.61 & 3.66 \\
pn-QRPA (C) & 0.70 & 4.00 & 9.54 & 8.99 & 2.47 & 2.74\\
PM (A) & 0.82 & 5.63 & 7.58 & 11.42 & 4.02 & 6.08 \\
PM (B) & 0.27  & 8.16 & 11.82 & 7.83 & 10.78 & 5.37  \\
SM (A)& 0.29 & 7.38 & 16.86 & 11.56 & 10.7 & 10.82 \\
SM (B)& 0.25 & 7.95 & 15.83 & 11.88 & 9.19 & 10.54 \\
SM (C) & 2.09 & 12.15  & 7.02 & 8.18 & 4.08 & 3.03 \\
$\mathbf{^{42}}$\bf{Ti} &  & &  &  &  &   \\
\tableline
pn-QRPA (A) & 1.25 & 3.36 & 5.92 & 5.63 & 3.08 & 3.22 \\
pn-QRPA (B) & 1.18 & 3.32 & 5.36 & 5.15 & 2.95 & 3.14 \\
pn-QRPA (C) & 2.64 & 4.76 & 3.47 & 4.26 & 2.50 & 2.88 \\
PM (A) & 0.97 & 5.86 & 12.18 & 11.21 & 4.08 & 2.01 \\
PM (B) & 1.42  & 7.42   & 4.57  &3.57   & 4.75  & 3.31  \\
SM (A)& 0.75 & 6.87 & 15.65 & 13.22 & 7.34 & 5.58 \\
SM (B)& 1.04 & 7.17 & 18.59 & 16.09 & 9.13 & 7.82\\
SM (C) &1.63   &7.30   &7.73   &8.14   &4.88   &3.45  \\
 $\mathbf{^{43}}$\bf{Ti} &  & &  &  &  &   \\
 \tableline
pn-QRPA (A) & 1.17 & 1.75 & 10.54 & 8.58 & 3.10 & 3.86\\
pn-QRPA (B) & 1.14 & 2.59 & 10.02 & 8.29 & 2.98 & 3.71\\
pn-QRPA (C) & 1.18 & 2.29 & 9.87 & 9.01 & 2.99 & 3.62\\
PM (A) &0.13& 0.83 &17.36   &5.41   &6.46  &7.11\\
PM (B) & 0.74 & 0.66 & 13.86 & 10.47 & 7.43 & 4.46 \\
SM (A)& 0.5 & 7.02 & 18.87 & 10.56 & 7.19 & 7.40 \\
SM (B)& 0.37 & 7.31 & 19.73 & 10.67 & 8.07 & 7.24\\
SM (C) & 2.37  & 6.06  & 5.06  & 7.27  &2.23   &2.81  \\
$\mathbf{^{44}}$\bf{Ti} &  & &  &  &  &   \\
\tableline
pn-QRPA (A) & 2.60 & 2.60 & 5.04 & 5.13 & 3.54 & 3.61 \\
pn-QRPA (B) & 1.79 & 1.76 & 6.37 & 6.59 & 1.90 & 1.85 \\
pn-QRPA (C) & 3.19 & 3.18 & 4.62 & 4.85 & 3.13 & 3.24 \\
PM (A) & 0.14 & 0.08 & 16.25 & 8.18 & 6.46 & 9.55  \\
PM (B) & 0.39 & 0.37  & 12.24 & 6.86 & 8.45 & 7.88 \\
SM (A)& 3.84 & 3.86 & 5.61 & 5.72 & 3.78 & 3.72 \\
SM (B)& 3.31 & 3.34 & 5.78 & 5.78 & 3.81 & 3.48\\
SM (C)  & 3.87 &  3.92 & 7.94 & 7.89 & 3.87 & 3.99 \\
\tableline
\end{tabular}
\tablenotetext{a}{pn-QRPA (A) are results of spherical pn-QRPA with
only ph channel} \tablenotetext{b}{pn-QRPA (B) are results of
spherical pn-QRPA with both ph+pp channels}
\tablenotetext{c}{pn-QRPA (C) are results of deformed pn-QRPA with
both ph+pp channels} \tablenotetext{d}{PM (A) are results of
spherical PM with only ph channel} \tablenotetext{e}{PM (B) are
results of spherical PM with both ph+pp channels}
\tablenotetext{f}{SMM (A) are results of spherical (SM) with both
ph+pp channels} \tablenotetext{g}{SM (B) are results of spherical SM
with both ph+pp channels}\tablenotetext{h}{SM(C) are results of
deformed SMM with both ph+pp channels}
\end{table}
\begin{table}
\small \caption{Same as Table~\ref{tbl-1} but for $^{45-50}$Ti.}
\label{tbl-2}
\begin{tabular}{ccccccc}
 $\mathbf{^{45}}$\bf{Ti} & $\mathbf{\sum B(GT_{-})}$ & $\mathbf{\sum B(GT_{+})}$
& $\mathbf{\bar{E}_{-}}$ & $\mathbf{\bar{E}_{+}}$ & $\mathbf{Width_{-}}$ & $\mathbf{Width_{+}}$  \\
\tableline
pn-QRPA (A) & 3.52 & 1.25 & 5.35 & 10.14 & 4.25 & 2.93\\
pn-QRPA (B) & 3.46 & 1.25 & 5.25 & 9.81 & 4.10 & 2.80\\
pn-QRPA (C) & 2.59 & 1.50 & 7.61 & 8.83 & 3.75 & 3.07\\
PM (A) & 0.39  & 0.39 &1.67   & 1.82 &9.69 &3.81  \\
PM (B) & 0.43  & 0.21 &  2.77 & 2.03  &  9.35 & 4.78 \\
SM (A)& 1.23 & 1.01 & 7.28 & 9.16 & 4.31 & 5.42 \\
SM (B)& 0.93 & 0.45 & 7.17 & 9.54 & 4.49 & 6.38\\
SM (C) & 2.72  & 2.1 &  8.49 & 8.58 &2.63 & 1.64\\
 $\mathbf{^{46}}$\bf{Ti} &  & &  &  &  &   \\
 \tableline
pn-QRPA (A) & 4.37 & 2.07 &4.66 & 5.04 & 3.41 & 3.10 \\
pn-QRPA (B) & 2.92 & 1.59 & 6.24 & 5.78 & 1.62 & 1.87 \\
pn-QRPA (C) & 4.48 & 2.31 & 4.82 & 4.11 & 2.95 & 3.09 \\
PM (A) & 6.81 & 1.16 & 5.71 & 0.44 & 3.72 & 2.35 \\
PM (B) & 6.55 & 0.43 & 6.05 & 6.44 & 2.81 & 6.82 \\
SM (A)& 9.20 & 3.55 & 8.04 & 6.14 & 6.34 & 5.04 \\
SM (B)& 6.64 & 2.97 & 6.25 & 5.01 & 4.05 & 3.64\\
SM (C)  & 8.60 &  2.94 & 10.00 & 7.42 & 3.25 & 3.93\\
$\mathbf{^{47}}$\bf{Ti} &  & &  &  &  &   \\
\tableline
pn-QRPA (A) & 3.72 & 0.40 & 8.08 & 8.08 & 3.65 & 3.20\\
pn-QRPA (B) & 3.64 & 0.39 & 7.73 & 7.70 & 3.44 & 2.97\\
pn-QRPA (C) & 3.63 & 0.39 & 7.75 & 7.70 & 3.44 & 2.98\\
PM (A) & 0.34 & 1.23 & 2.07 & 1.38 & 7.96 & 4.22 \\
PM (B) & 0.38 & 1.36 & 3.58 & 3.04 & 8.13 & 4.18\\
SM (A)& 2.79 & 0.59 & 12.03 & 11.35 & 8.08 & 8.27 \\
SM (B)& 2.61 & 0.44 & 9.55 & 10.19 & 4.67 & 6.96\\
SM (C) & 9.16 & 1.44   &8.93   & 6.38  & 1.89   & 1.45  \\
$\mathbf{^{48}}$\bf{Ti} &  & &  &  &  &   \\
\tableline
pn-QRPA (A) & 6.19 & 1.84 & 4.38 & 4.21 & 3.25 & 2.72 \\
pn-QRPA (B) & 4.12 & 1.58 & 6.10 & 4.31 & 1.44 & 2.11 \\
pn-QRPA (C) & 6.12 & 1.80 & 5.69 & 3.57 & 3.00 & 2.94 \\
PM (A) & 11.71 & 0.005 & 2.08 & 8.96 & 0.49 & 5.45 \\
PM (B) & 11.85 & 0.29 & 2.40 & 3.08 & 1.62 & 5.62\\
SM (A)& 14.10 & 2.10 & 7.28 & 5.83 & 3.36 & 1.92 \\
SM (B)& 13.86 & 1.84 & 7.23 & 5.57 & 3.40 & 1.05\\
SM (C)  & 13.85 & 2.09 & 12.76 & 6.42 & 3.55 & 4.19 \\
 $\mathbf{^{49}}$\bf{Ti} &  & &  &  &  &   \\
 \tableline
pn-QRPA (A) & 3.81 & 0.31 & 7.36 & 6.66 & 5.06 & 3.69\\
pn-QRPA (B) & 4.07 & 0.31 & 7.20 & 6.39 & 4.88 & 3.46\\
pn-QRPA (C) & 5.59 & 0.29 & 7.80 & 6.72 & 3.23 & 2.78\\
PM (A) &6.93  &0.02  & 10.59 & 11.27 & 1.06 &7.35 \\
PM (B) &5.29  &0.18  &10.08  &9.13  &1.59  &6.42 \\
SM (A)& 4.22 & 0.50 & 9.09 & 8.02 & 4.02 & 3.91 \\
SM (B)& 5.17 & 0.44 & 8.75 & 7.56 & 4.04 & 3.65 \\
SM (C) &16.00   & 1.43  &   12.33& 8.06  &2.97   &1.73  \\
 $\mathbf{^{50}}$\bf{Ti} &  & &  &  &  &   \\
 \tableline
pn-QRPA (A) & 8.10 & 1.64 & 4.55 & 3.88 & 3.40 & 3.23 \\
pn-QRPA (B) & 8.03 & 1.56 & 4.28 & 3.41 & 3.28 & 3.04 \\
pn-QRPA (C) & 8.01 & 1.57 & 5.80 & 3.86 & 3.10 & 3.44 \\
PM (A) & 18.75 & 0.83 & 5.65 & 1.47 & 2.19 & 1.90\\
PM (B) & 18.60 & 0.67 & 5.59 & 1.96 & 2.35 & 4.16 \\
SM (A)& 19.66 & 1.73 & 8.28 & 2.40 & 3.53 & 1.95 \\
SM (B)& 20.01 & 1.59 & 8.37 & 2.16 & 3.48 & 1.99\\
SM (C)  & 19.13 & 1.78 & 14.69 & 6.45 & 3.89 & 4.38 \\
 \tableline
\end{tabular}
\end{table}
\begin{table}
\small \caption{Same as Table~\ref{tbl-1} but for $^{51-56}$Ti.}
\label{tbl-3}
\begin{tabular}{ccccccc}
 $\mathbf{^{51}}$\bf{Ti} & $\mathbf{\sum B(GT_{-})}$ & $\mathbf{\sum B(GT_{+})}$
& $\mathbf{\bar{E}_{-}}$ & $\mathbf{\bar{E}_{+}}$ & $\mathbf{Width_{-}}$ & $\mathbf{Width_{+}}$  \\
\tableline
pn-QRPA (A) & 8.78 & 0.27 & 8.10 & 7.62 & 2.93 & 0.79\\
pn-QRPA (B) & 8.83 & 0.27 & 8.07 & 7.43 & 2.92 & 0.78\\
pn-QRPA (C) & 7.90 & 0.27 & 8.35 & 7.29 & 2.87 & 1.00\\
PM (A) & 4.99 & 1.11 &9.42  &4.80  &1.72  &1.62 \\
PM (B) & 4.67 & 1.10 & 9.42 & 4.34 & 1.75 & 2.02\\
SM (A)& 6.44 & 0.92 & 11.62 & 5.56 & 4.24 & 1.99 \\
SM (B)& 7.08 & 0.83 & 13.31 & 5.19 & 7.62 & 1.97 \\
SM (C) & 24.61  & 0.30   &13.77   &9.51   &4.89   &6.66 \\
$\mathbf{^{52}}$\bf{Ti} &  & &  &  &  &   \\
\tableline
pn-QRPA (A) & 10.08 & 1.52 & 6.19 & 3.96 & 3.25 & 4.37 \\
pn-QRPA (B) & 10.02 & 1.43 & 5.96 & 3.33 & 3.12 & 3.89 \\
pn-QRPA (C) & 10.01 & 1.41 & 6.13 & 3.31 & 3.12 & 3.63\\
PM (A) & 24.67 & 0.88 & 8.20 & 4.64 & 2.25 & 2.61 \\
PM (B) & 23.45 & 0.24 & 7.94 & 11.45 & 2.74 & 6.94  \\
SM (A)& 25.40 & 1.61 & 14.69 & 5.22 & 3.73 & 4.43 \\
SM (B)& 25.21 & 1.41 & 14.43 & 4.17 & 3.53 & 2.18 \\
SM (C) & 24.73 & 1.69 & 16.55 & 5.82 & 3.88 & 4.22 \\
 $\mathbf{^{53}}$\bf{Ti} &  & &  &  &  &   \\
 \tableline
pn-QRPA (A) & 9.35 & 0.14 & 8.89 & 6.44 & 3.20 & 2.74 \\
pn-QRPA (B) & 11.09 & 0.23 & 7.00 & 4.08 & 3.48 & 2.16 \\
pn-QRPA (C) & 10.00 & 0.25 & 7.28 & 4.50 & 3.70 & 2.73 \\
PM (A) & 10.55  &0.43  &8.41  &5.03  &1.99  &2.38 \\
PM (B) & 11.89 &  0.09&  7.71& 13.09 &4.19  &7.76 \\
SM (A)& 25.40 & 1.61 & 14.69 & 5.22 & 3.73 & 4.43 \\
SM (B)& 1.96 & 0.66 & 9.05 & 4.57 & 5.59 &2.18 \\
SM (C) &38.71  &0.44   &13.05   &4.61   &3.43   &3.14  \\
 $\mathbf{^{54}}$\bf{Ti} &  & &  &  &  &   \\
 \tableline
pn-QRPA (A) & 12.08 & 1.40 & 6.53 & 2.77 & 3.20 & 4.56 \\
pn-QRPA (B) & 12.03 & 1.29 & 6.26 & 2.14 & 3.11 & 4.03 \\
pn-QRPA (C) & 11.96 & 1.26 & 6.74 & 2.43 & 3.16 & 4.00 \\
PM (A) & 29.93 & 0.77 & 7.43 & 6.16 & 2.71 & 3.27 \\
PM (B) & 29.00 & 0.29 & 7.43 & 11.54 & 2.86 & 6.70 \\
SM (A)& 30.99 & 1.44 & 16.7 & 4.85 & 3.91 & 5.43 \\
SM (B)& 3.81 & 1.24 & 16.44 & 3.54 & 3.65 & 2.64\\
SM (C)  & 35.14 & 0.27 & 20.40 & 8.87 & 4.00 & 6.55 \\
$\mathbf{^{55}}$\bf{Ti} &  & &  &  &  &   \\
 \tableline
pn-QRPA (A) & 9.50 & 0.13 & 10.19 & 5.65 & 3.27 & 2.92 \\
pn-QRPA (B) & 12.56 & 0.14 & 8.05 & 6.93 & 4.40 & 2.49 \\
pn-QRPA (C) & 11.35 & 0.22 & 7.75 & 4.04 & 6.62 & 4.06 \\
PM (A) & 10.08 & 0.26 & 9.83 & 8.42 & 2.93 & 3.27 \\
PM (B) &11.16  &0.10  &7.70  &11.77  &2.98  &6.98 \\
SM (A)& 17.62 & 0.84 & 17.72 & 5.16 & 3.47 & 3.64 \\
SM (B)& 18.21 & 0.77 & 17.89 & 4.52 & 2.96 & 2.17\\
SM (C)  &41.96   &0.18   &13.24  &6.01   &4.27   & 3.25 \\
$\mathbf{^{56}}$\bf{Ti} &  & &  &  &  &   \\
\tableline
pn-QRPA (A) & 13.98 & 0.88 & 7.06 & 2.13 & 3.43 & 4.86 \\
pn-QRPA (B) & 13.10 & 0.11 & 7.30 & 11.16 & 2.93 & 15.08 \\
pn-QRPA (C) & 13.49 & 0.54 & 7.52 & 2.85 & 3.26 & 4.74 \\
PM (A) & 36.89 & 1.69 & 6.94 & 2.29 & 2.64 & 3.05  \\
PM (B) & 35.95 & 0.23 & 6.93 & 11.69 & 2.66 & 5.39 \\
SM (A)& 36.32& 1.03 & 18.22 & 1.92 & 4.04 & 2.76 \\
SM (B)& 36.17 & 0.92 & 18.14 & 3.10 & 4.01 &7.13 \\
SM (C) & 34.45 & 0.84 & 19.67 & 5.72 & 4.49 & 5.55 \\
\tableline
\end{tabular}
\end{table}
\begin{table}
\small \caption{Same as Table~\ref{tbl-1} but for $^{57-60}$Ti.}
\label{tbl-4}
\begin{tabular}{ccccccc}
 $\mathbf{^{57}}$\bf{Ti} & $\mathbf{\sum B(GT_{-})}$ & $\mathbf{\sum B(GT_{+})}$
& $\mathbf{\bar{E}_{-}}$ & $\mathbf{\bar{E}_{+}}$ & $\mathbf{Width_{-}}$ & $\mathbf{Width_{+}}$  \\
 \tableline
pn-QRPA (A) & 14.13 & 0.11 & 10.34 & 4.95 & 3.59 & 3.66\\
pn-QRPA (B) & 13.92 & 0.06 & 10.10 & 8.95 & 3.29 & 3.00\\
pn-QRPA (C) & 14.23 & 0.20 & 7.61 & 2.33 & 5.12 & 3.70\\
PM (A) & 14.15 &0.91  &10.99  &2.68  &8.62  &2.97 \\
PM(B) & 20.78 & 0.12 & 7.79 & 11.86 & 5.54 & 5.22 \\
SM (A)& 11.5 & 0.55 & 18.64 & 2.34 & 4.51 & 2.29 \\
SM (B)& 14.40 & 0.49 & 19.29 & 3.00 & 5.04 & 5.76\\
SM (C)  & 47.48  & 0.16  &15.14   & 6.64  & 4.44  &3.59  \\
$\mathbf{^{58}}$\bf{Ti} &  & &  &  &  &   \\
\tableline
pn-QRPA (A) & 15.88 & 0.71 & 7.34 & 2.55 & 3.81 & 5.46 \\
pn-QRPA (B) & 14.14 & 0.10 & 7.91 & 11.85 & 3.01 & 4.90 \\
pn-QRPA (C) & 15.65 & 0.50 & 7.49 & 2.55 & 3.71 & 5.02 \\
PM (A) & 42.09 & 1.05 & 6.26 & 3.39 & 3.39 & 4.28\\
PM (B) & 39.29 & 0.31 & 6.59 & 10.76 & 3.23 & 4.63 \\
SM (A)& 41.47 & 0.41 & 18.93 & 2.72 & 4.72 & 4.08 \\
SM (B)& 41.33 & 0.24 & 18.86 & 6.14 & 4.71 &7.41 \\
SM (C) & 39.05 & 0.57 & 20.55 & 6.92 & 4.93 & 6.09 \\
 $\mathbf{^{59}}$\bf{Ti} &  & &  &  &  &   \\
 \tableline
pn-QRPA (A) & 15.34 & 0.06 & 10.36 & 5.32 & 3.87 & 5.02\\
pn-QRPA (B) & 14.64 & 0.02 & 10.64 & 9.13 & 3.26 & 3.33\\
pn-QRPA (C) & 16.70 & 0.15 & 7.23 & 1.54 & 5.59 & 3.17\\
PM (A) & 16.49  &0.55  &8.17  &2.93  &6.89  &3.98 \\
PM (B) &23.73  &0.12  &6.67  &11.39  &4.31  &4.73 \\
SM (A)& 14.17 & 0.22 & 18.54 & 2.23 & 5.01 & 3.08 \\
SM (B)& 7.87 & 0.12 & 19.08 & 4.53 & 8.82 & 6.65\\
SM (C) & 35.81  & 0.07 &  15.62 &  7.93 &  5.04 &3.21  \\
 $\mathbf{^{60}}$\bf{Ti} &  & &  &  &  &   \\
 \tableline
pn-QRPA (A) & 17.70 & 0.50 & 7.28 & 3.07 & 4.03 & 5.75\\
pn-QRPA (B) & 15.27 & 0.09 & 8.15 & 11.86 & 3.03 & 4.35 \\
pn-QRPA (C) & 17.67 & 0.39 & 7.34 & 2.85 & 3.98 & 5.35 \\
PM (A) & 47.59 & 0.66 & 5.58 & 6.20 & 4.19 & 4.65 \\
PM (B) & 46.53 & 0.71 & 5.49 & 6.86 & 4.06 & 5.83 \\
SM (A)& 47.03 & 0.15 & 19.29 & 5.65 & 5.49 & 5.52 \\
SM (B)& 46.92 & 0.09 & 19.16 & 8.76 & 5.53 & 6.28\\
SM (C)  & 40.69 & 0.46 & 20.01 & 8.28 & 5.31 & 6.05 \\
\tableline
\end{tabular}
\end{table}
\begin{table}
\small \caption{Total GT$_{-}$ strengths and calculated centroids
for all GT distribution functions of $^{46}$Ti shown in
Fig.~\ref{fig3}} \label{tbl-5}
\begin{tabular}{cccc}
 \bf{Panel} & $\mathbf{\sum B(GT_{-})}$ & $\mathbf{\bar{E}_{-}}$ (MeV) &
 Cutoff energy in daughter (MeV) \\
\tableline
Exp.  & 1.77 & 2.53 & 4.38\\
pn-QRPA (C) & 4.00 & 4.31 & 14.84\\
SM (C) & 8.32 & 9.66 & 15.99\\
PM (B) & 6.43 & 9.87 & 15.06\\
LSSM & 4.17 & 5.56 & 15.21\\
SM:KB3G int. & 2.25 & 2.76 & 5.76\\
SM:GXPF1 int. & 2.03 & 2.80 & 5.96\\
QD & 1.74 & 2.83 & 5.42 \\
\end{tabular}
\end{table}
\end{document}